\documentclass[aps,prd,twocolumn,superscriptaddress,eqsecnum,nofootinbib,showpacs]{revtex4}
\usepackage{graphicx}
\usepackage{euscript,amssymb}
\newcommand{\uno}{{\text{\scshape i}}}
\newcommand{\dos}{{\text{\scshape ii}}}
\newcommand{\tres}{{\text{\scshape iii}}}
\newcommand{\fuera}{{\text{\scshape a}}}
\newcommand{\dentro}{{\text{\scshape f}}}

\begin{document}
\title{Quantum behavior of FRW radiation-filled universes}
\author{Mariam Bouhmadi-L\'{o}pez}
 \email{mbouhmadi@imaff.cfmac.csic.es}
\affiliation{ Instituto de Matem\'{a}ticas y F\'{\i}sica Fundamental,\\
Consejo Superior de Investigaciones Cient\'{\i}ficas,\\ C/ Serrano 121,
28006 Madrid, Spain}
\author{Luis J. Garay}
 \email{garay@imaff.cfmac.csic.es}
\affiliation{ Instituto de Matem\'{a}ticas y F\'{\i}sica Fundamental,\\
Consejo Superior de Investigaciones Cient\'{\i}ficas,\\ C/ Serrano 121,
28006 Madrid, Spain}
\author{Pedro F. Gonz\'{a}lez-D\'{\i}az}
 \email{p.gonzalezdiaz@imaff.cfmac.csic.es}
\affiliation{ Instituto de Matem\'{a}ticas y F\'{\i}sica Fundamental,\\
Consejo Superior de Investigaciones Cient\'{\i}ficas,\\ C/ Serrano 121,
28006 Madrid, Spain}
\date{April 10, 2002}
\begin{abstract}
We study the quantum vacuum fluctuations around closed
Friedmann-Robertson-Walker (FRW) radiation-filled universes with
nonvanishing cosmological constant. These vacuum fluctuations are
represented by a conformally coupled massive scalar field and are
treated in the lowest order of perturbation theory. In the
semiclassical approximation, the perturbations are governed by
differential equations which, properly linearized, become
generalized Lam\'{e} equations. The wave function thus obtained must
satisfy appropriate regularity conditions which ensure its
finiteness for every field configuration. We apply these results
to asymptotically anti de-Sitter Euclidean wormhole spacetimes and
show that there is no catastrophic particle creation in the
Euclidean region, which would lead to divergences of the wave
function.
\end{abstract}


\maketitle

\section{Introduction}

Homogeneity and isotropy of the universe on large scale is a good
approximation to describe the classical behavior of the universe.
Friedmann-Robertson-Walker (FRW) models are specially designed to
implement these properties. Nevertheless, seeds of inhomogeneity
and anisotropy are needed in order to describe the cosmic
structure. For this purpose, studies of cosmological perturbations
are necessary. Seminal works in this direction were done in
Ref.~\cite{Lifshitz} and later on in Ref.~\cite{Ginsparg} where
the authors studied the stability of de Sitter space.

The classical description of the universe breaks down for energies
of the order or above the Planck scale. Therefore, it is necessary
to use a quantum theory of gravity and to postulate some boundary
conditions for the universe in order to describe its initial
state. Despite the absence of a fully consistent quantum theory of
gravity, many studies have been carried out that shed light on the
problem of the creation of the universe with different boundary
conditions \cite{Hawking1,Linde,Vilenkin3,hawking,wada,Vilenkin2}.
These works characterize the quantum behavior of the universe in
the semiclassical approximation through its wave function in both
minisuperspace and superspace, where the inhomogeneous and
anisotropic modes are included perturbatively in the models.

In Ref.~\cite{Rubakov2}, it was noted that the wave function of a
closed FRW universe with a positive cosmological constant becomes
infinite in the forbidden (tunneling) region when the universe is
filled with radiation and subject to vacuum fluctuations of a
massive scalar field conformally coupled to gravity. In other
words, the author concluded that during the tunneling process a
catastrophic particle creation takes place. He also speculated
that perhaps these phenomena might be a rather common feature of
tunneling processes due to quantum gravity effects. In
Ref.~\cite{wada}, it was shown that the wave function of a de
Sitter universe in the presence of gravitational perturbations
increase for some boundary conditions but never diverge. Similar
results were obtained in Ref.~\cite{Vilenkin2} for a minimally
coupled scalar field and tunneling boundary conditions.

In this paper, we develop a method based on Ref.~\cite{Vilenkin2}
to study the quantum behavior of the wave function of a
radiation-filled FRW universe with cosmological constant and
radiation, which includes vacuum fluctuations represented by a
massive scalar field conformally coupled to gravity. These vacuum
fluctuations will be regarded as perturbations to the homogeneous
and isotropic solutions of the Wheeler-DeWitt equation. We can
deduce, at least for some values of the scalar field mass and
negative cosmological constant, that the perturbed wave function
is not divergent in the classically forbidden regime. As we will
see, the finiteness of the wave function is due to the regularity
and boundary conditions, which although restrictive still allow
for finite solutions. These quantum states represent
asymptotically anti de Sitter wormholes \cite{Barcelo1,Barcelo3}.

The paper is organized as follows. In Sec.~\ref{classical}, we
review the classical behavior of a closed radiation-filled FRW
universe with a cosmological constant, both in the Lorentzian and
Euclidean regions. In Sec.~\ref{III}, we derive the Wheeler-Dewitt
equation for these universes in the presence of vacuum
fluctuations of a conformally coupled massive scalar field and
perform the semiclassical approximation. In Sec.~\ref{IV}, we
deduce the general matching conditions that relate the wave
function defined in the different semiclassical regions. We also
impose the regularity conditions. In Sec.~\ref{V}, we obtain the
background wave function and linearize the equations for the
matter vacuum fluctuations thus obtaining generalized Lam\'{e}
equations. We solve these equations for asymptotically anti-de
Sitter wormhole spacetimes. We show that this perturbed wave
function is finite for all possible values of the scale factor and
scalar field configurations. Finally in Sec.~\ref{VI} we summarize
our results and conclude.

\section{Lorentzian and Euclidean behavior of FRW universes}
\label{classical}

The main part of this paper will be devoted to study the quantum
behavior of a closed FRW universe filled with radiation against
perturbations due to a massive scalar field conformally coupled to
gravity. But before that, let us shortly review the classical
behavior of a closed FRW universe filled with radiation
\cite{Halliwell}. In our analysis we include a cosmological
constant $\Lambda\equiv3 \lambda$, and we represent, for
simplicity, the radiation of our universe by a conformal scalar
field $\tilde{A}$. The FRW metric can be written as
\[
ds^2=a(\bar \eta)^2(-d\bar\eta^2+d\Omega_3^2) ,
\]
where $\bar \eta$ is the Lorentzian conformal time and
$d\Omega_3^2$ is the line element on the unit three-sphere.
Writing the radiation field as
\[
\tilde{A}({\mathbf{x}},\bar\eta)=\frac{1}{\sqrt{2\pi}}\tilde{\chi}(\bar\eta)/a(\bar\eta),
\]
the Lorentzian equation of motion for $\tilde{\chi}(\bar\eta)$
becomes
\begin{eqnarray}
&\tilde{\chi}''+\tilde{\chi}=0,& \label{campohomo}
\end{eqnarray}
where the prime denotes derivative with respect to $\bar \eta$.
The equation for this field can be integrated to obtain a constant
of motion $\tilde{K}$, related to the energy density by
$\rho=\tilde{K}/a^4$:
\[
\tilde{\chi}'^2+\tilde{\chi}^2=\frac{3\tilde{K}}{2\textrm{G}},
\]
where $\textrm{G}$ is the gravitational constant. Then the scale
factor must satisfy the equation
\begin{equation}
a'^2 +V(a)-\tilde{K} = 0, \label{factorescala}
\end{equation}
where
\begin{equation}
V(a)=a^2-\lambda a^4. \label{potencial}
\end{equation}
\begin{figure}
\includegraphics[width=\columnwidth]{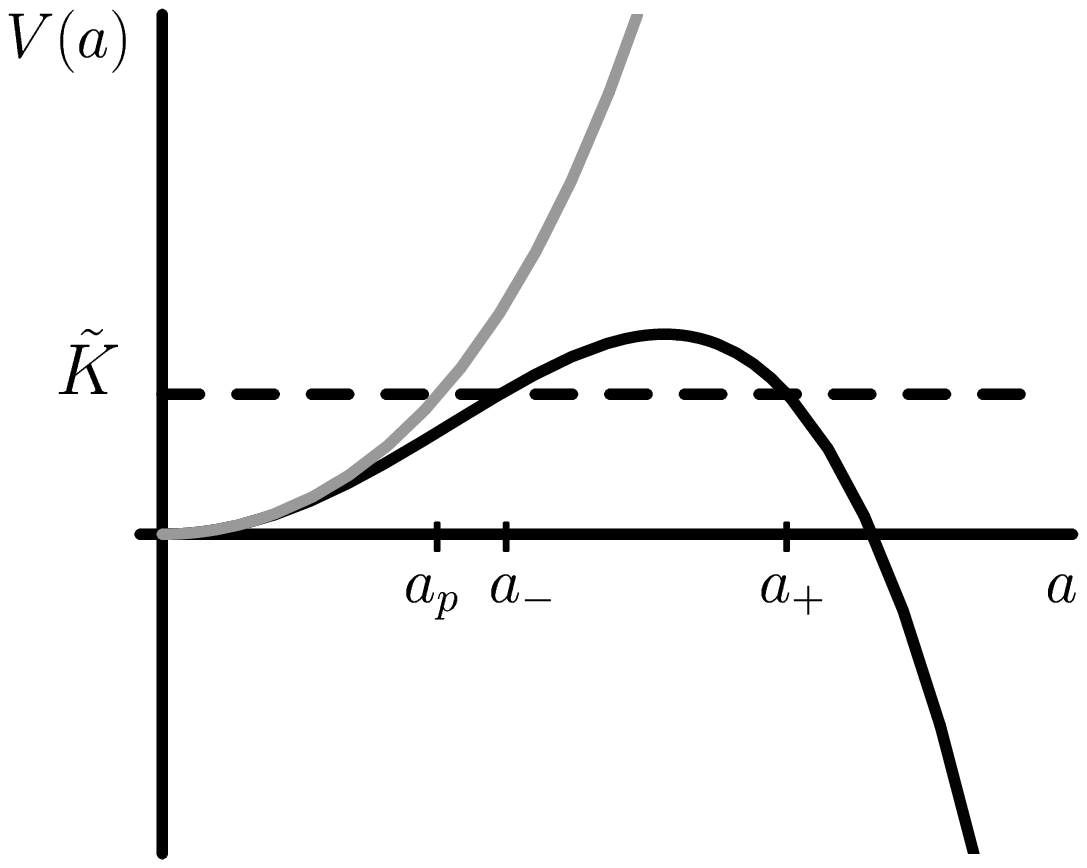}
\caption{This figure shows the potencial $V(a)$ defined on
Eq.~(\ref{potencial}). The darkest and lightest curves correspond
to a positive and negative cosmological constant
$\Lambda=3\lambda$ cases, respectively. The parameter $\tilde{K}$
is related to the amount of radiation presents in the FRW
universe. On the one hand, for positive $\lambda$ and $\tilde{K}$
smaller than the maximum of $V(a)$, as in the case plotted, $a_-$
represents the maximum radius of the collapsing radiation-filled
FRW universe, while $a_+$ represents the minimum scale factor of
the asymptotically de Sitter universe. On the other hand, for
negative $\lambda$ the scale factor $a_p$ represents the maximum
radius of the collapsing universe.} \label{graficapotencial}
\end{figure}
The shape of the potential $V(a)$ depends on the sign of the
cosmological constant $\lambda$. For a positive cosmological
constant, it increases up to a maximum value $\frac{1}{4\lambda}$
at $a=\frac{1}{\sqrt{2\lambda}}$, and decreases after that for
scale factors larger than $\frac{1}{\sqrt{2\lambda}}$
Fig.~(\ref{graficapotencial}). For a negative cosmological
constant the situation is rather different, as the potential is
always increasing and never negative
Fig.~(\ref{graficapotencial}).

We can distinguish three kinds of behavior for $a$. The first one
describes a collapsing universe. This is the case when the
cosmological constant is negative and $\tilde{K}\neq0$, for which
\begin{equation}
a(\bar\eta)=a_p {\rm
cn}\left[\sqrt{m}(\bar\eta-\bar\eta_{+}),\frac{m-1}{2m}\right],
\label{afuera}\end{equation} where
$\bar\eta\in[\bar\eta_-,\bar\eta_+]$,
$a_p^2=\frac{1-m}{2\lambda}$, $m=\sqrt{1-4\tilde{K}\lambda}$ and
$\sqrt{m}(\bar\eta_+-\bar\eta_-)={\rm K}(\frac{m-1}{2m})$. In
these expressions, ${\rm cn}[x,\frac{m-1}{2m}]$ is a Jacobian
Elliptic function and ${\rm K}(\frac{m-1}{2m})$ is the complete
Jacobian elliptic integral or quarter-period function
\cite{Abramowitz,Gradshteyn}. Note that $\bar\eta_+$ is an
arbitrary constant that can be set equal to zero. The scale factor
of this universe increase from $a=0$ at $\bar\eta=\bar\eta_-$ up
to $a=a_p$ for $\bar\eta=0$, which is the maximum radius of this
universe. The other case of a collapsing closed FRW universe
corresponds to a positive cosmological constant and a value of the
parameter $\tilde{K}$, related to the amount of radiation present
in the universe, smaller than the maximum of the potential $V(a)$,
i.e., $0<\tilde{K}<\frac{1}{4\lambda}$. Under these conditions,
the scale factor in terms of the cosmological time, $d\bar
t=a(\bar\eta)d\bar\eta$, has the expression
\[
a(\bar
t)^2=\frac{1}{2\lambda}\left\{1-m\cosh\left[2\sqrt{\lambda}(\bar
t-\bar t_-)\right]\right\},
\]
the maximum value of the scale factor is
$a_-^2=\frac{1-m}{2\lambda}$, the corresponding to $\bar t =\bar
t_-$ which is the solution of the algebraic equation
$V(a)=\tilde{K}$. For both solutions, the maximum radius of the
universe increases with the amount of radiation, given by
$\tilde{K}$.

The second kind of solutions describes an asymptotically de Sitter
space-time when $\lambda>0$, whose scale factor is given by
\begin{eqnarray}
a(\bar t)^2& =
&\frac{1}{2\lambda}\left\{\sqrt{4\tilde{K}\lambda}\sinh\left[2\sqrt{\lambda}(\bar
t-\bar t_+)\right]-\right.\nonumber\\ & -
&\left.\cosh\left[2\sqrt{\lambda}(\bar t -\bar
t_+)\right]+1\right\}\nonumber
\end{eqnarray}
for $\tilde{K}>\frac{1}{4\lambda}$, and
\[
a(\bar
t)^2=\frac{1}{2\lambda}\left\{1+m\cosh\left[2\sqrt{\lambda}(\bar
t-\bar t_+)\right]\right\},
\]
for $\tilde{K}<\frac{1}{4\lambda}$, where $\bar t\in[\bar
t_+,+\infty)$. The difference between these two cases is that for
sufficient radiation, i.e. $\tilde{K}$ is larger than the maximum
of the potential $V(a)$, the scale factor grows from zero up to
infinity, while in the opposite case; i.e. for $\tilde{K}$ smaller
than the maximum of the potential $V(a)$, the scale factor grows
from a minimum value different from zero,
$a_+^2=\frac{1+m}{2\lambda}$, due to the presence of the barrier
of potential $V(a)$, to become asymptotically de Sitter.

And finally, there is a third kind of solutions which exactly
coincides with a de Sitter space-time
\[
a(\bar t)^2=\frac{1}{\lambda}\cosh^2\left[\sqrt{\lambda}(\bar
t-\bar t_+)\right],
\]
in the absence of radiation and for a positive $\lambda$.

It can be checked that there are not classical solutions of the
Einstein equations corresponding to a closed FRW in absence of
radiation, $\tilde{K}=0$, with a negative cosmological constant.
It is only for $\tilde{K}>0$ that it is possible to have a
Lorentzian evolution for the scale factor $a$.

Up to now, we have described the different possible Lorentzian
solutions for a closed homogeneous and isotropic universe filled
with radiation and we have seen that the potential $V(a)$ forbids
the classical evolution for some values of the scale factor.
Therefore two classical FRW universes filled with radiation with
$\tilde{K}<\frac{1}{4\lambda}$ are disconnected and the scale
factor for a FRW universe with a negative cosmological constant
has a maximum value when the content of the universe corresponds
to radiation.

As it is well known, the fact that two classically allowed
universes separated by a potential barrier are classically
disconnected does not mean that they cannot be connected quantum
mechanically. In the lowest approximation, this connection is
established by an instanton whose explicit form can be obtained by
performing an analytical continuation of Eq.~(\ref{factorescala}),
for a positive $\lambda$, so that the classically forbidden region
is now the permitted one. The solution for the scale factor must
satisfy
\[
a(\eta_-)=a_-,\qquad a(\eta_+)=a_+,
\]
in order to connect with the two classical FRW universes. From the
analytically continued version of Eq.~(\ref{factorescala}), we
obtain the following solution for the scale factor:
\begin{equation}
a(\eta)^2 = \frac{1+m}{2\lambda}\; {\rm
dn}^2\left[\sqrt{\frac{1+m}{2}}(\eta-\eta_+),\frac{2m}{1+m}\right],
\label{instanton}
\end{equation}
where $\eta\in [\eta_-,\eta_+]$ with
$\eta_+-\eta_-=-\sqrt{\frac{2}{1+m}}{\rm K}(\frac{2m}{1+m})$. In
this expression ${\rm dn}[x,\frac{2m}{1+m}]$ is the Jacobian
elliptic delta-amplitude function \cite{Abramowitz,Gradshteyn}.
Note that $\eta_+$ is an arbitrary constant that can be set equal
to zero. This instanton was also found in Ref. \cite{Halliwell},
where the authors considered a closed FRW with a material content
corresponding to a massless scalar field conformally coupled to
gravity. In the absence of radiation, $m=1$, the turning points of
the potential $V(a)$, i.e., the solution of $V(a)=\tilde{K}$, (see
Eq.~(\ref{potencial})) becomes $a_-=0$, $a_+=\sqrt{1/\lambda}$.
The instanton (\ref{instanton}) then acquires the simple form
\[
a(\eta)^2=\frac{1}{\lambda} {\rm sech}^2(\eta-\eta_+)
\]
and $\eta_+-\eta_-=\infty$.

While for a positive cosmological constant the Euclidean solution
for a closed FRW universe filled with radiation connects two
classical solutions, for a negative $\lambda$ the solution behaves
as an Euclidean asymptotically anti de Sitter wormhole. This can
be easily deduced from the analytical continuation of
Eq.~(\ref{factorescala}) to imaginary conformal time. The scale
factor of the wormhole looks like~\cite{Barcelo1}
\begin{equation}
a(\eta)^2=a_p^2{\rm
nc}^2\left[\sqrt{m}(\eta-\eta_-),\frac{1+m}{2m}\right],
\label{wormhole}
\end{equation}
where $\eta\in[\eta_-,\eta_+]$ and $\sqrt{m}(\eta_+-\eta_-)={\rm
K}(\frac{1+m}{2m})$. In this expression ${\rm
nc}[x,\frac{1+m}{2m}]$ is a Jacobian Elliptic function
\cite{Abramowitz,Gradshteyn}. The value $a_p$ describes the radius
of the wormhole throat.

\section{The wave function of the universe}
\label{III}

The quantum behavior of the FRW universe can be described by the
solution to the Wheeler-DeWitt equation \cite{Wheeler}. In the WBK
approximation, the wave functions can be approximated, under
certain conditions, by ingoing and outgoing modes defined through
the classical action in the Lorentzian section, while in the
Euclidean sector, the wave function can be approximated by linear
combinations of increasing and decreasing modes in terms of the
Euclidean action. Boundary conditions that determine these linear
combinations are also necessary. In this section, we will obtain
the general shape of the wave function of a closed FRW universe
filled with radiation and whose content corresponds to a massive
scalar field conformally coupled to gravity.

\subsection{Canonical formulation}

We will consider a minisuperspace described by two degrees of
freedom, the scale factor $a$ and a homogeneous and isotropic
scalar field conformally coupled to gravity $\tilde{A}$. Around
this minisuperspace, we will study the linear perturbations due to
an inhomogeneous and anisotropic massive scalar field $\Phi$
conformally coupled to gravity. We will obtain the Wheeler-DeWitt
equation from a specific representation of the Hamiltonian of the
system which can be constructed easily from the classical action
of the system
\begin{eqnarray}
S & = &S_{{\rm g}}+S_{\rm r}+S_{\rm m}, \nonumber\\ S_{{\rm g}}& =
& {\frac{1}{16\pi{\rm G}}\int
d^4x\sqrt{-g}\left(R-6\lambda\right)} - \nonumber \\ & - &
{\frac{1}{8\pi{\rm G}} \int d^3x\sqrt{h}K}, \nonumber \\ S_{{\rm
r}}& = &{\int d^4x\sqrt{-g}\left[-\frac{1}{2}(\nabla \tilde{A})^2-
\frac{1}{12}R \tilde{A}^2\right]}+ \nonumber \\ & + &
{\frac{1}{6}\int d^3x\sqrt{h}K \tilde{A}^2}, \nonumber \\ S_{{\rm
m}}& = &{\int d^4x\sqrt{-g}\left[-\frac{1}{2}\nabla\Phi^2-
\frac{1}{2}\left(\mu^2+\frac{1}{6}R\right)\Phi^2\right]}+\nonumber
\\ & + &\frac{1}{6}\int d^3x\sqrt{h}K\Phi^2,
\label{accion}
\end{eqnarray}
where ${\rm G}$ is the gravitational constant, $\mu$ is the mass
of the scalar field and $K$ is the trace of the extrinsic
curvature. We have used the sign of conventions of Misner, Thorne
and Wheeler \cite{Misner}.

We introduce new variables which correspond to an expansion in
hyperspherical harmonics of the massive scalar field around the
background solution described in the previous section as follows
\begin{eqnarray}
\Phi({\mathbf{x}},\bar\eta)& =
&a(\bar\eta)^{-1}\bigg[\varphi(\bar\eta)+
\sum_{nlm}f_{nlm}(\bar\eta)Q^{nlm}({\mathbf{x}})\bigg]\label{campo}\\
\tilde{A}({\mathbf{x}},\bar\eta)&=&\frac{1}{\sqrt{2\pi}}a(\bar\eta)^{-1}\bigg[\tilde\chi(\bar\eta)+
\sum_{nlm}\zeta_{nlm}(\bar\eta)Q^{nlm}({\mathbf{x}})\bigg]\nonumber\\
\end{eqnarray}
where $Q^{nlm}$ are the scalar hyperspherical harmonics,
eigenfunctions of the 3-dimensional Laplacian $\nabla^2$ in the
three-sphere, i.e., they satisfy the eigenvalue equation $\nabla^2
Q^{nlm}=-n(n+2)Q^{nlm}$, with $n=0,1,2,\dots$. The mode $n=0$
corresponds to the homogeneous and isotropic perturbation, while
higher values of $n$ correspond to inhomogeneous and anisotropic
modes. We have considered the background solution of the massive
scalar field $\Phi$, equal to zero, i.e. $\varphi=0$. From now on
and for the sake of simplicity, we will drop the indices $lm$ and
keep only the eigenvalue index $n$ in all the expressions.

The space-time metric must also be expanded around the homogeneous
and isotropic background solution. If we write the Lorentzian
metric in the $3+1$ form
\[
ds^2=a^2(-N^2+N^i N_i)d\eta^2+2aN_i d\eta dx^i + g_{ij} dx^i dx^j,
\]
this expansion can be written as
\begin{eqnarray}
N&=&1+\sum_n g_nQ^n, \nonumber\\ N_i&=&\sum_n (j_n P^n_i +k_n
S^n_i), \nonumber\\ g_{ij}&=&a^2(\Omega_{ij} +\epsilon_{ij}),
\nonumber\\ \epsilon_{ij}&=&\sum_n (a_nQ^n_{ij}+b_nP^n_{ij} +c_n
S^n_{ij}+d_nG^n_{ij}),\nonumber
\end{eqnarray}
$\Omega_{ij}$ being the metric in the unit three-sphere and
${Q^n}$, ${P^n}$, ${S^n}$ and ${G^n}$ are the standard
hyperspherical harmonics on the three-sphere
\cite{Lifshitz,hawking}.

The coefficients $x_n\equiv\{a_n,b_n,c_n,g_n,j_n,k_n\}$ can in
principle be eliminated by means of a diffeomorphism on the
three-sphere and choosing suitable lapse and shift functions
\cite{hawking,wada}. The only terms in these expansions that
cannot be gauged away correspond to pure transverse traceless
tensor perturbations that describe gravitational waves. These are
represented by the coefficients $d_n$. The Lorentzian action up to
second order in perturbations has the form
\[
S=S^{(0)}[a,\chi]+S_{g+r}^{(2)}[a,\chi,\zeta_n,x_n,d_n]+S_m^{(2)}[a,f_n].
\]
This can be seen from Eq.~(\ref{accion}) taking into account that
the massive background field vanishes; i.e. $\varphi=0$. Indeed in
this case the perturbations of the massive scalar field decouple
from perturbations of the metric and the radiation up to second
order. Since we are interested in the behavior of closed FRW
universes against the quantum fluctuations of the vacuum of the
massive scalar field $\Phi$, we have chosen its homogeneous
background mode to vanish. So the explicit expression of
$S_{g+r}^{(2)}$ is not necessary to study the behavior of the FRW
universe under perturbations of the massive scalar field. The zero
order action $S^{(0)}$ and the second order action of the scalar
field perturbations $S_m^{(2)}$ have the form
\begin{eqnarray}
S^{(0)}[a,\chi]& = &
\int_{\bar\eta_-}^{0}d\bar\eta\left[\frac{3\pi}{4\rm{G}}
(a^2-a'^2-\lambda
a^4)+\frac{\pi}{2}(\tilde\chi'^2-\tilde\chi^2)\right],\nonumber
\\ S^{(2)}_{\rm m}[a,f_n]& = &\sum_{n}\int_{\bar\eta}^{0}
d\bar\eta \left(\frac{1}{2}f_{n}'^2-\frac{1}{2}U_n(\bar\eta)
f_n^2\right),
\end{eqnarray}
where
\[
U_n(\bar\eta)=(n+1)^2+\mu^2 a(\bar\eta)^2.
\]
Finally, the Hamiltonian of the system can be written as
\begin{eqnarray}
H & = &H^{(0)}[a,\chi]+ H^{(2)}_{\rm m}[a,f_n],\nonumber \\
H^{(0)}[a,\chi]& = &
-\frac{\rm{G}}{3\pi}\;P_a^2-\frac{3\pi}{4\rm{G}}a^2+
\frac{3\pi}{4\rm{G}}\lambda
a^4\;+\;\frac{1}{2}(P_{\tilde\chi}^2+\tilde\chi^2), \nonumber
\\
H^{(2)}_{\rm m}[a,f_n]& = &
\sum_{n}\frac{1}{2}P_{f_{n}}^2+\frac{1}{2}U_n(\bar\eta) f_n^2.
\label{hamiltoniano}
\end{eqnarray}

This Hamiltonian describes the classical constraint of our system
$H=0$ and is related with the invariance of the Lorentzian action
under time reparametrizations. This constraint will be our
starting point for studying the quantum behavior of closed FRW
universes filled with radiation.

The constraint $H=0$ in the context of quantum gravity becomes a
constraint on the wave function of the universe leading to the
Wheeler-DeWitt equation, which can be written as
\begin{eqnarray}
&\displaystyle\left\{-\frac{\textrm{G}}{3\pi}\partial^2_a+
\frac{3\pi}{4\textrm{G}}(a^2-\lambda a^4)+\frac{1}{2}
(\partial^2_{\tilde\chi}-\tilde\chi^2)+\right.\nonumber&
\\
&\left.\displaystyle{\frac{1}{2}
\sum_{n=1}^{+\infty}\left\{\partial^2_{f_n}-U_n(a)f_n^2\right\}
}\right\}\widetilde{\Psi}(a,f_n,\tilde\chi)=0&, \label{ww1}
\end{eqnarray}
where
\[
U_n(a)=(n+1)^2+\mu^2a^2.
\]
The functional dependence of the wave function on the radiation
field can be obtained by separation of variables. The part of the
wave function $\Psi(a,f_n)$ which depends on the other degrees of
freedom present in our model, $a$ and $f_n$, must satisfy
\begin{equation}
\left[-\frac{\textrm{G}}{3\pi}\partial^2_a+
\frac{3\pi}{4\textrm{G}}V_0(a)+
\sum_{n=1}^{+\infty}(\frac{1}{2}\partial^2_{f_n} -V_n(a)f_n^2)
\right]\Psi(a,f_n)=0, \label{ww}
\end{equation}
where the potentials $V_n(a)$ and $V_0(a)$ are defined as follows:
\begin{eqnarray}
V_n(a)&=&\frac{1}{2}[(n+1)^2+\mu^2 a^2],\;
n=1,2,\ldots,\nonumber\\ V_0(a)&=&a^2-\lambda a^4-\tilde{K}.
\label{redefpotencial}
\end{eqnarray}
Here $\tilde{K}$ is a separation constant, related to the energy
of the mode $\tilde\chi$ and quantifies the amount of radiation
present in the universe as we have seen in Sec.~\ref{classical}.

\subsection{Semiclassical approximation}
\label{semiclassical}

We will treat the quantum behavior of radiation-filled FRW in the
semiclassical approximation \cite{wada,Vilenkin2}, where the
physical lengths involved in our problem are larger than Planck
length $l_p$. In this approximation, the solutions to equation
(\ref{ww}) will be written as linear combinations of
\begin{eqnarray}
\Psi_1(a,f_n)=\exp\left[-\frac{1}{\textrm{G}}S_0(a)-
\frac{1}{2}\sum_{n=1}^{+\infty} S_{n}(a)f_n^2\right],\nonumber\\
\Psi_2(a,f_n)=\exp\left[-\frac{1}{\textrm{G}}R_0(a)-
\frac{1}{2}\sum_{n=1}^{+\infty} R_{n}(a)f_n^2\right],
\label{debajo}
\end{eqnarray}
where $\Psi_1$ and $\Psi_2$ will be decreasing and increasing
functions in the classically forbidden regime or ingoing and
outgoing waves in the classically allowed regime \cite{Vilenkin2}.
We will deal firstly with the regions of validity of the
semiclassical approximation in term of the amount of radiation
present in the universe and in later stage of our study, we will
specify clearly the kind of functions on each regime, they will
depend on the inclusion of the Lorentzian or the Euclidean time.
$S_0$ and $R_0$ are related to the unperturbed part of the wave
function of the universe while $S_n$ and $R_n$ are related to the
vacuum fluctuations of the massive scalar field and corresponds to
the perturbative part of the wave function of the universe. Using
the WBK approximation, we obtain
\begin{eqnarray}
\left(\frac{dS_0}{da}\right)^2=\frac{9\pi^2}{4}V_0(a),
\label{sn1}\\
\frac{dS_0}{da}\frac{dS_n}{da}-\frac{3\pi}{2}S_n^2=-3\pi V_n.
\label{sn2}
\end{eqnarray}
where we have performed an asymptotic expansion on G and likewise
for $R_0$ and $R_n$.
 Similarly to
the usual WBK this approximation is valid as long as
\begin{equation}
\textrm{G}|\frac{dV_0(a)}{da}|\ll|V_0(a)|^{\frac{3}{2}}\label{valwbk}.
\end{equation}
From now on, we will consider that the amount of radiation present
on the closed FRW universes in the case of positive cosmological
constant $\lambda$, is such that this kind of models has two
classically disconnected solutions, i.e. $0<\tilde{K}<1/4\lambda$,
allowing for quantum tunelling between the two universes. As can
be seen from the validity of the WBK approximation
Eq.~(\ref{valwbk}), this condition breaks down near the points
$a_{-}$ and $a_{+}$ in the case of $\lambda>0$. These points
corresponds respectively to the maximum radius of the collapsing
universe and the minimum radius of the asymptotically de Sitter
universe. Similarly the WBK method is not valid near the point
$a_p$ which corresponds to the maximum radius of the collapsing
universe in the case of a negative $\lambda$. Therefore we have to
analyze the behavior of the wave function near these turning
points using other methods as explained below.

\subsection{Turning points}
\label{turning points}

Starting from the expansion of the potentials $V_0(a)$ and
$V_n(a)$ around each turning point, $a_-$, $a_+$ and $a_p$, the
Wheeler-DeWitt equation (\ref{ww}) acquires the form
\begin{eqnarray}
\left\{-\frac{\textrm{G}}{3\pi}\partial^2_a+
\frac{3\pi}{4\textrm{G}}\frac{dV_0(a)}{da}|_{a=a_i}(a-a_i)\right.\nonumber\\
\left. + \sum_{n=1}^{+\infty}\left[\frac{1}{2}\partial^2_{f_n}
-V_n(a_i)f_n^2\right]\right\}\Psi(a,f_n)=0, \label{wwl}
\end{eqnarray}
where $a_i= a_{\pm},a_p$. This linear approximation hold if and only
if:
\begin{equation}
|(a-a_i)\frac{d^2V_0(a)}{d^2a}|_{a=a_i}|\ll2|\frac{dV_0(a)}{da}|_{a=a_i}.
\label{vallineal}
\end{equation}

The wave functions of the closed FRW universes can be expressed as
linear combinations of $\Psi_1(a,f_n)$ and $\Psi_2(a,f_n)$ defined
in Eq.~(\ref{debajo}), where $S_0$, $R_0$, $S_n$ and $R_n$ satisfy
the differential equations (\ref{sn1},\ref{sn2}), if the value of
the scale factor, $a$, is such that the condition (\ref{valwbk})
holds. While in the linear regime the behavior of the wave
functions of these universes will be solution of Eq.~(\ref{wwl}).
As we will see there are values of the radius of the universe
where both conditions Eqs.~(\ref{valwbk},\ref{vallineal}) hold and
it is possible to connect the wave function on the WBK regimes
through the linear regime.

Let us begin analyzing the case of a positive cosmological
constant, in which the conditions for the linear approximation are
\begin{eqnarray}
|a-a_+|\ll\frac{2m}{2+3m}a_+, \;
|a-a_-|\ll|\frac{2m}{-2+3m}|\;a_-\nonumber\\
\end{eqnarray}
respectively around  $a_+$ and $a_-$, while near these turning points
the WBK conditions (\ref{valwbk}) read:
\begin{eqnarray}
\textrm{G}^{2/3}(2a_+m)^{-1/3}\ll|a-a_+|,\nonumber\\
\textrm{G}^{2/3}(2a_-m)^{-1/3}\ll|a-a_-|,
\end{eqnarray}
where the first one corresponds to $a_+$ and the second one to
$a_-$. So it is a sufficient condition that
\begin{eqnarray}
(\textrm{G}\lambda)^2\ll|\frac{4m^4(1+m)^2}{(2+3m)^3}|,\nonumber\\
(\textrm{G}\lambda)^2\ll|\frac{4m^4(1-m)^2}{(-2+3m)^3}|,
\end{eqnarray}
to conclude the existence of values of $a$ such that there exists
an overlapping between the WBK and the linear approximations. This
overlapping depends only on the amount of radiation present in the
FRW universes when it can be described semiclassically
$\lambda\ll\textrm{G}^{-1}$.

In the case of a negative cosmological constant $\lambda<0$, we
have to deal with a unique turning point $a_p$. Using similar
methods to the case $\lambda>0$, we obtain that a sufficient
condition for the existence of value of $a$ such that the linear
and WBK approximation hold is:
\begin{equation}
(\textrm{G}\lambda)^2\ll\frac{4m^4(m-1)^2}{(3m-2)^3},
\end{equation}
which is the case when the maximum value of the radius of the
collapsing closed FRW universe, $a_p$, is large enough
(equivalently the parameter $m$ is large).

\section{Matching conditions}
\label{IV}

Using the fact that there are values of the scale factor, $a$, in
which the linear and WBK approximations hold, we will connect the
wave function on the different WBK regimes through the linear ones
around the turning points, $a_-$, $a_+$ and $a_p$.

\subsection{Positive cosmological constant}

For the case of a positive cosmological constant there are three
WBK regimes, one corresponding to values of the scale factor such
that $a<a_-$ and the wave function, $\Psi_{\uno}$, can be written
as
\begin{eqnarray}
\Psi_{\uno}(a,f_n)& = &C\exp\left[-\frac{1}{\textrm{G}}
S^{\uno}_0(a)- \frac{1}{2}\sum_{n=1}^{+\infty}
S^{\uno}_{n}(a)f_n^2\right]+\nonumber\\ & +
&D\exp\left[-\frac{1}{\textrm{G}} R^{\uno}_0(a)-
\frac{1}{2}\sum_{n=1}^{+\infty}
R^{\uno}_{n}(a)f_n^2\right];\nonumber\\ \label{funcion1}
\end{eqnarray}
a second one, which is classically forbidden, $a_-<a<a_+$, and for
which the wave function, $\Psi_{\dos}$, can be expressed as
follows
\begin{eqnarray}
\Psi_{\dos}(a,f_n)& =
&A\exp\left[-\frac{1}{\textrm{G}}S^{\dos}_0(a)-
\frac{1}{2}\sum_{n=1}^{+\infty}
S^{\dos}_{n}(a)f_n^2\right]+\nonumber\\ & + &
B\exp\left[-\frac{1}{\textrm{G}} R^{\dos}_0(a)-
\frac{1}{2}\sum_{n=1}^{+\infty}
R^{\dos}_{n}(a)f_n^2\right];\nonumber \\ \label{funcion2}
\end{eqnarray}
and finally, a third one in which the scale factor, $a$, is larger
than $a_+$. In this regime the wave function is
\begin{eqnarray}
\Psi_{\tres}(a,f_n)& =
&E\exp\left[-\frac{1}{\textrm{G}}S^{\tres}_0(a)-
\frac{1}{2}\sum_{n=1}^{+\infty}
S^{\tres}_{n}(a)f_n^2\right]+\nonumber\\ & + &
F\exp\left[-\frac{1}{\textrm{G}}R^{\tres}_0(a)-
\frac{1}{2}\sum_{n=1}^{+\infty} R^{\tres}_{n}(a)f_n^2\right].
\nonumber \\ \label{funcion3}
\end{eqnarray}
In all these expressions, the functions with the subscript $0$
represent the background part of the wave function and satisfy the
differential equation (\ref{sn1}), while the functions with suffix
$n$, $n\geq1$, correspond to the perturbations of the wave
function of the universe and they are solutions of the
differential equation stated in expression (\ref{sn2}). Outside
the potential barrier $V(a)$, the functions $S_0^{\uno}$ and
$S_0^{\tres}$ are related to the outgoing modes and the functions
$R_0^{\uno}$ and $R_0^{\tres}$ correspond to the ingoing modes. On
the other hand, under the potential barrier $S_0^{\dos}$ and
$R_0^{\dos}$ represent growing and decreasing background terms,
respectively, of the wave function $\Psi_{\dos}$.

To connect the wave function of the FRW universe
Eqs.~(\ref{funcion1},\ref{funcion2},\ref{funcion3}) through the
linear regimes around $a_-$ and $a_+$ we will consider that we
have a unique mode $f_n$ of the massive scalar field $\Phi$. The
general case can be easily deduced from this one. Near the turning
points, the wave function of the closed FRW universe filled with
radiation and a positive cosmological constant in addition to the
vacuum fluctuation of a massive conformally coupled scalar field,
can be expressed as
\begin{eqnarray}
\Psi_i(a,f_n)&=&\sum_{m_{i}=0}^{+\infty}\left[\gamma_{m_i}Ai(z_i)+\delta_{m_i}Bi(z_i)\right]\times
\nonumber\\ & \times &\exp(-y_{n_i}^2/2){\rm H}_{m_i}(y_{n_i}),
\label{linealizada}
\end{eqnarray}
where
\begin{eqnarray}
z_i& =
&(-1)^{i+1}\left(\frac{3\pi}{2\textrm{G}}\right)^{2/3}\left[(-1)^{i+1}
\partial_a{V}_0(a_{\mp})\right]^{1/3}(a-a_{\mp})+\nonumber\\ & + &
2\left(\frac{2\textrm{G}}{3\pi}\right)^{1/3}
\left[(-1)^{i+1}\partial_a{V}_0(a_{\mp})\right]^{-2/3}\beta_{n_i},\nonumber\\
\beta_{n_i}& = &-(2m_i+1)\sqrt{V_n(a_{\pm})/2}; \;
m_i\in\mathbb{N},\nonumber\\ y_{n_i}& =
&\left(2V_n(a_{\pm})\right)^{1/4}f_n. \label{defbeta}
\end{eqnarray}
in this expressions the index $i$ may takes the values 1 or 2
which correspond to the wave function on the linear regimes around
$a_-$ and $a_+$ respectively, the potential $V_0(a)$ was defined
in Eq.~(\ref{redefpotencial}), $\gamma_{m_i}$ and $\delta_{m_i}$
are constants, the functions $Ai(x)$ and $Bi(x)$ are the Airy
functions and ${\rm H}_{m_i}(x)$ are the Hermite's polynomials
\cite{Abramowitz}.

Let us now begin connecting the WBK wave function under the
barrier, $\Psi_{\dos}(a,f_n)$, around $a_-$ with the linear regime
using the fact that there are values of the scale factor $a$, such
that both approximations, the WBK and the linear one, hold as was
explained in Sec.~\ref{turning points}. For these values of the
scale factor, the wave function
 $\Psi_{\dos}(a,f_n)$, using Eqs.~(\ref{sn1},\ref{funcion2}), can be expressed near $a_-$ as
\begin{eqnarray}
\Psi_{\dos}(a,f_n)& = &
A\exp\left[\frac{\pi}{\textrm{G}}\sqrt{\partial_a{V}_0(a_-)}\;
(a-a_-)^{3/2}\right]\times\nonumber \\ & \times
&\exp\left(-\frac{1}{2}S^{\dos}_n(a_)f_n^2\right)+ \nonumber \\ &
+ &
B\exp\left[-\frac{\pi}{\textrm{G}}\sqrt{\partial_a{V}_0(a_-)}\;(a-a_-)^{3/2}\right]\times\nonumber
\\ & \times & \exp\left(-\frac{1}{2}R^{\dos}_n(a_)f_n^2\right).
\label{funa-derecha}
\end{eqnarray}
Let us do some remarks about the behavior of the variables $z_i$.
These variables are related to the different indices $m_i$ of the
Hermite's polynomials, through the constants $\beta_{n_i}$
Eq.~(\ref{defbeta}). Consequently, the functions $Ai(z_i)$ and
$Bi(z_i)$ can not be factor out from the sums over the different
indices $m_i$ which define the wave functions $\Psi_i$ near the
turning points. However, since we are working on the semiclassical
framework the second term, which is proportional to $\beta_{n_i}$,
on the definition of $z_i$ is much smaller than the first one, as
the first one is proportional to $\textrm{G}^{-2/3}$ while the
second one is proportional to $\textrm{G}^{1/3}$. So the variables
$z_i$ in this regime effectively do not depend on the indices
$m_i$ and the Airy functions can be factor out from the sum given
in Eq.~(\ref{linealizada}). Considering our last statement, we
will connect the wave function on the linear regime with that of
the WBK regime for values of $z_i$ such that the Airy functions
can be approximated by their asymptotic behaviors
($z_i\rightarrow\pm\infty$)\footnote{ To use the asymptotic
behavior of the Airy functions on the expression
(\ref{linealizada}), it is necessary to check that the condition
$|z_i|\rightarrow+\infty$, the linear and the WBK approximation
overlap for some values of the scale factor, $a$, near the turning
points. Indeed this is the case because the condition
$|z_i|\rightarrow+\infty$ and the validity of the WBK
approximation (\ref{valwbk}) coincide near all the turning points
$a_-$, $a_+$ and $a_p$, in particular for the quantities of
radiation that we are considering. \label{1}}.

Near the turning point $a_-$, the variable $z_1$ is positive under
the barrier and reaches values large enough to use the asymptotic
behavior of the Airy functions. Once the $Ai(z_1)$ and $Bi(z_1)$
have been substituted by their asymptotic behavior in expression
(\ref{linealizada}) for $i=1$, comparing the resulting expression
for the wave function of the universe $\Psi_1(a,f_n)$ near the
turning points $a_-$ for $a > a_-$ with the wave function
$\Psi_{\dos}(a,f_n)$ in Eq.~(\ref{funa-derecha}), and imposing the
continuity of the FRW wave function, we see that the growing term
in $\Psi_{\dos}$, related to $S^{\dos}_0$, overlaps with the
$Bi(z_1)$ terms on the expression of $\Psi_1$, while the
decreasing term in $\Psi_{\dos}$, related to $R^{\dos}_0$,
overlaps with the $Ai(z_1)$ terms on the expression of $\Psi_1$.
Also we obtain the following equations
\begin{eqnarray}
\sum_{m_1=0}^{+\infty}\gamma_{m_1}\exp\left(-\frac{y^2_{n_1}}{2}\right)
H_{m_1}(y_{n_1}) = \nonumber \\ B\exp\left
[-\frac{1}{2}R^{\dos}_n(a_-)f^2_n \right ], \nonumber\\
\sum_{m_1=0}^{+\infty}\delta_{m_1}\exp\left
(-\frac{y^2_{n_1}}{2}\right) H_{m_1}(y_{n_1}) = \nonumber \\
A\exp\left [-\frac{1}{2}S^{\dos}_n(a_-)f^2_n\right ]. \\
\label{gammaizquierda}
\end{eqnarray}
Using the orthogonality relations of the Hermite polynomials
\cite{Abramowitz} and the following formula \cite{Gradshteyn}:
\begin{equation}
\int_{-\infty}^{+\infty}\exp(-x^2)H_{2m}(xy)dx=\sqrt{\pi}
\frac{(2m)!}{m!}(y^2-1)^m, \label{integral}
\end{equation}
we have
\begin{displaymath}
\gamma_{m_1}=\left\{ \begin{array}{ll} B(1-k_-)^{l_1}/(2^{2l_1}
{l_1}!k_{-}^{l_1+1/2}), \; m_1=2l_1, \; l_1\in\mathbb{N}
\\
0, \qquad \qquad \qquad \qquad \qquad \; m_1=2l_1+1, \;
l_1\in\mathbb{N},
\end{array}\right.
\end{displaymath}
\begin{displaymath}
\delta_{m_1}=\left\{ \begin{array}{ll}
A(1-\tilde{k}_-)^{l_1}/(2^{2l_1}
{l_1}!{\tilde{k}_{-}}^{l_1+1/2}),\; m_1 = 2l_1,\; l_1\in\mathbb{N}
\\
0, \qquad \qquad \qquad \qquad \qquad \; m_1 = 2l_1+1, \;
l_1\in\mathbb{N},
\end{array}\right.
\end{displaymath}
where
\begin{equation}
2k_-=\frac{R^{\dos}_n(a_-)}{\sqrt{2V_n(a_-)}}+1, \;
2\tilde{k}_-=\frac{S^{\dos}_n(a_-)}{\sqrt{2V_n(a_-)}}+1.\nonumber
\end{equation}
The last expressions determine the linear behavior of the wave
function near the point $a_-$ in terms of the WBK wave function
under the barrier $V(a)$ (see Eq.~(\ref{potencial})) which can be
explicitly seen through the dependence of the coefficients
$\delta_{m_1}$ and $\gamma_{m_1}$ in $A$ and $B$ respectively.

Now, using on the one hand the explicit expression of
$\Psi_1(a,f_n)$ around $a_-$ for values of the variable $z_1$ such
that $z_1\rightarrow-\infty$ and on the another hand the behavior
of the wave function $\Psi_{\uno}$ of the FRW universe outside the
potential, $V(a)$ (see Eq.~(\ref{potencial})),
$\Psi_{\uno}(a,f_n)$ near the turning point $a_-$, we can obtain
similar relations to the ones expressed in
Eq.~(\ref{gammaizquierda}) for the coefficients $\gamma_{m_1}$,
$\delta_{m_1}$, $C$ and $D$:
\begin{eqnarray}
\sum_{m_1=0}^{+\infty}\left
(\gamma_{m_1}/(2i)-\delta_{m_1}/2\right)
\exp(-i\frac{\pi}{4})\exp\left(-\frac{y^2_{n_1}}{2}\right)\times\nonumber\\
 \times H_{m_1}(y_{n_1}) = C\exp\left
[-\frac{1}{2}S^{\uno}_n(a_-)f^2_n \right ], \nonumber\\
\sum_{m_1=0}^{+\infty}\left(\delta_{m_1}/2 +
\gamma_{m_1}/(2i)\right) \exp(i\frac{\pi}{4})\exp\left
(-\frac{y^2_{n_1}}{2}\right)\times\nonumber\\ \times
H_{m_1}(y_{n_1}) = D\exp\left
[-\frac{1}{2}R^{\uno}_n(a_-)f^2_n\right ], \nonumber \\
\label{gammaderecha}
\end{eqnarray}
where we have used the continuity of the wave function of the FRW
universe. Finally using expressions
(\ref{gammaizquierda},\ref{gammaderecha}), we deduce
\begin{eqnarray}
C& = &\left[A/2-B/(2i)\right]\exp\left(-i\frac{\pi}{4}\right),
\nonumber \\ D& = &
\left[A/2+B/(2i)\right]\exp\left(i\frac{\pi}{4}\right), \nonumber
\\
S^{\uno}_n(a_-)& = &
S^{\dos}_n(a_-)=R^{\uno}_n(a_-)=R^{\dos}_n(a_-),
\label{matchinga-}
\end{eqnarray}
which determine the wave function of a FRW universe filled with
radiation outside the potential barrier $V(a)$, in terms of the
wave function of a FRW universe inside the potential barrier
$V(a)$ and viceversa.

Once we have obtained the matching conditions for the wave
function in the case of a positive cosmological constant for the
turning point $a_-$, let us deal with the matching conditions for
the turning point $a_+$. For this purpose, we approximate the wave
function, under the potential barrier $V(a)$, $\Psi_{\dos}(a,
f_n)$ defined in Eq.~(\ref{funcion2}) near the point $a_+$ by
\begin{eqnarray}
& & \Psi_{\dos}(a,f_n)=\nonumber\\ & & =
A\exp\bigg[\frac{3\pi}{2\textrm{G}}\int_{a_-}^{a_+}\sqrt{V_0(a)}da
-\frac{\pi}{\textrm{G}}\sqrt{-\partial_a{V}_0(a_+)}\times
\bigg.\nonumber\\ & &\times \bigg.(a_+\; -a)^{3/2}\bigg]
\exp\left(-\frac{1}{2}S^{\dos}_n(a_+)f_n^2\right)+\nonumber\\ & &
+
B\exp\bigg[-\frac{3\pi}{2\textrm{G}}\int_{a_-}^{a_+}\sqrt{V_0(a)}da
+\frac{\pi}{\textrm{G}}\sqrt{-\partial_a{V}_0(a_+)}\times\bigg.\nonumber
\\ & & \times  \bigg.(a_+\; -a)^{3/2}\bigg]
\exp\left(-\frac{1}{2}R^{\dos}_n(a_+)f_n^2\right).
\label{funa+izquierda}
\end{eqnarray}

In the next step, we will match the wave function $\Psi_{\dos}(a,
f_n)$, using the last equation, with the wave function on the
linear regime around $a_+$, $\Psi_2(a, f_n)$ expressed in
Eq.~(\ref{linealizada}) for $i=2$. Similarly to our procedure for
the matching conditions in $a_-$, we use the asymptotic behavior
of the Airy's functions in the expression of $\Psi_2(a, f_n)$ for
values of the scale factor, $a$, such that the condition
$z_2\rightarrow+\infty$, the WBK approximation and the linear one
hold and we obtain the following relations between the
coefficients $\gamma_{m_2}$, $\delta_{m_2}$ and the constants $A$
and $B$
\begin{eqnarray}
&&\sum_{m_2=0}^{+\infty}\gamma_{m_2}\exp\left(-\frac{y^2_{n_2}}{2}\right)
H_{m_2}(y_{n_2})= \\ &&
A\exp\left[\frac{3\pi}{2\textrm{G}}\int_{a_-}^{a_+}\sqrt{V_0(a)}da\right]
\exp\left [-\frac{1}{2}S^{\dos}_n(a_+)f^2_n \right ], \nonumber\\
&&\sum_{m_1=0}^{+\infty}\delta_{m_2}\exp\left
(-\frac{y^2_{n_2}}{2}\right) H_{m_2}(y_{n_2})= \nonumber \\ &&
B\exp\left[-\frac{3\pi}{2\textrm{G}}\int_{a_-}^{a_+}\sqrt{V_0(a)}da\right]
\exp\left [-\frac{1}{2}R^{\dos}_n(a_+)f^2_n\right ].\nonumber
\label{gamma+izquierda}
\end{eqnarray}

Using as before the orthogonality relations of the Hermite
polynomials, we have
\begin{displaymath}
\gamma_{m_2}=\left\{ \begin{array}{ll} A\Delta
(1-k_+)^{l_2}/(2^{2l_2} {l_2}!{ k_{+}}^{l_2+1/2}), \; m_2=2l_2,
\;\\ 0, \qquad\qquad\qquad\qquad\qquad\qquad m_2=2l_2+1, \;
\end{array}\right.
\end{displaymath}
\begin{displaymath}
\delta_{m_2}=\left\{ \begin{array}{ll} B\Delta^{-1}
(1-\tilde{k}_+)^{l_2}/(2^{2l_2} {l_2}!{\tilde{k}_{+}}^{l_2+1/2}),
\; m_2=2l_2,\;
\\
0, \qquad\qquad\qquad\qquad\qquad\qquad m_2=2l_2+1,
\end{array}\right.
\end{displaymath}
where
\begin{eqnarray}
2k_+=\frac{S^{\dos}_n(a_+)}{\sqrt{2V_n(a_+)}}+1, \;
2\tilde{k}_+=\frac{R^{\dos}_n(a_+)}{\sqrt{2V_n(a_+)}}+1,\nonumber\\
\Delta=\exp\left[\frac{3\pi}{2\textrm{G}}\int_{a_-}^{a_+}\sqrt{V_0(a)}da\right],\;l_2\in\mathbb{N}.
\end{eqnarray}
These expressions define the behavior of the wave function
$\Psi_2(a,f_n)$ in the linear regime around the turning point
$a_+$ in term of the WBK wave function $\Psi_{\dos}(a,f_n)$ under
the potential barrier $V(a)$.

Using the asymptotic behavior of the wave function
$\Psi_2(a,f_n)$, this time for $z_2\rightarrow+\infty$, we can
match the wave function of the linear regime around $a_+$ with the
wave function $\Psi_{\tres}(a,f_n)$ outside the barrier of
potential~$V(a)$. The continuity of the wave function implies
\begin{eqnarray}
&&\sum_{m_2=0}^{+\infty}\left(-\gamma_{m_2}/(2i)+\delta_{m_2}/2\right)
\exp(-i\frac{\pi}{4})\exp\left(-\frac{y^2_{n_2}}{2}\right)\times
\nonumber \\ & \times & H_{m_2}(y_{n_2}) = E\exp\left
[-\frac{1}{2}S^{\tres}_n(a_+)f^2_n \right ] \nonumber\\
&&\sum_{m_2=0}^{+\infty}\left(\delta_{m_2}/2 +
\gamma_{m_2}/(2i)\right) \exp(i\frac{\pi}{4})\exp\left
(-\frac{y^2_{n_2}}{2}\right)\times \nonumber
\\ & \times &H_{m_2}(y_{n_2})
= F\exp\left [-\frac{1}{2}R^{\tres}_n(a_+)f^2_n\right
],\nonumber\\ \label{gamma+derecha}
\end{eqnarray}
and therefore
\begin{eqnarray}
E & = & \left[-\frac{A}{2i}\exp\left(\frac{3\pi}{2\textrm{G}}
\int_{a_-}^{a_+}\sqrt{V_0(a)}\right)\right.+ \nonumber
\\ & + & \left. \frac{B}{2}
\exp\left(-\frac{3\pi}{2\textrm{G}}\int_{a_-}^{a_+}\sqrt{V_0(a)}\right)\right]
\exp(-i\frac{\pi}{4}), \nonumber\\ F& =
&\left[\frac{A}{2i}\exp\left(\frac{3\pi}{2\textrm{G}}
\int_{a_-}^{a_+}\sqrt{V_0(a)}\right)+\right. \nonumber\\ & +
&\left.\frac{B}{2}
\exp\left(-\frac{3\pi}{2\textrm{G}}\int_{a_-}^{a_+}\sqrt{V_0(a)}\right)\right]
\exp(i\frac{\pi}{4}), \nonumber \\ &&S_n^{\dos}(a_+) =
S_n^{\tres}(a_+)=R_n^{\dos}(a_+)=R_n^{\tres}(a_+).
\label{matching+}
\end{eqnarray}
These equalities, together with Eq.~(\ref{matchinga-}) are the
matching conditions for the wave function of a FRW universe filled
with radiation, a positive cosmological constant and the vacuum
fluctuations of a massive conformally coupled scalar field. Apart
from these matching conditions, there are other conditions which
ensure the good behavior of the wave function. These are the
regularity conditions \cite{Vilenkin2}
\begin{eqnarray}
\mathit{Re}\left[S_n^{\uno}(a)\right],
\;\mathit{Re}\left[R_n^{\uno}(a)\right]>0,&& \; \textrm{for} \;
a<a_-,\nonumber\\ \mathit{Re}\left[S_n^{\dos}(a)\right],
\;\mathit{Re}\left[R_n^{\dos}(a)\right]>0,&& \; \textrm{for} \;
a_-<a<a_+,\nonumber\\ \mathit{Re}\left[S_n^{\tres}(a)\right],
\;\mathit{Re}\left[R_n^{\tres}(a)\right]>0,&& \; \textrm{for} \;
a_+<a. \label{regularity+}
\end{eqnarray}

\subsection{Tunneling boundary conditions of the universe}

As an example that illustrates the applications of these boundary
condition, let us discuss the tunneling boundary conditions of the
universe \cite{Vilenkin3}. For these boundary conditions, outside
and far from the potential barrier $V(a)$, i.e. in the classically
allowed region and for values of the scale factor much larger than
the turning point $a_+$, the wave function of the universe should
contain only outgoing modes. That is, the coefficient $F$ (see
Eq.~(\ref{funcion3})) must be equal to zero, so that no ingoing
modes appear on the asymptotically de Sitter region. Once the
tunneling boundary conditions have been applied, we can deduce the
linear combination of the growing and decaying terms that define
the FRW wave function under the barrier, i.e. the relationship
between the constants $A$ and $B$ (see Eq.~(\ref{matching+}))
\begin{equation}
|A/B|=\exp\left(-\frac{3\pi}{\textrm{G}}\int_{a_-}^{a_+}
\sqrt{V_0(a)}\right),
\end{equation}
in agreement with the results obtained in Ref.~\cite{Barcelo2} for
an analogous system.

So, both growing and decaying background terms are present in the
expression of $\Psi_{\dos}(a,f_n)$ under the barrier $V(a)$.
Nevertheless, we see that the growing term associated to
$S_0^{\dos}(a)$ is multiplied by the constant $A$ which is
exponentially smaller than the constant $B$, which multiply the
decreasing term on $\Psi_{\dos}(a,f_n)$. We see that, even if it
allows the appearance of a growing term on the classically
forbidden region, under the barrier, it is exponentially reduced.
On the other hand the wave function $\Psi_{\uno}(a,f_n)$ defined
on the classically allowed collapsing region $a<a_-$, will be a
combination a of ingoing and outgoing modes.

\subsection{Negative cosmological constant}
In the case of a negative cosmological constant, $\lambda<0$,
there is a unique turning point $a_p$. This value of the scale
factor separates a classically allowed region, $a<a_p$, from a
classically forbidden one, $a_p<a$. The matching conditions for
the wave function around $a_p$ can be deduced carrying out a
similar analysis to the one presented previously for $\lambda>0$.
We summarize our results in what follows.

In the classically allowed region ($a<a_p$), we will denote the
wave function by
\begin{eqnarray}
\Psi_{\fuera}(a,f_n)& = &A_{\fuera}\exp\left[-\frac{1}{\textrm{G}}
S^{\fuera}_0(a)- \frac{1}{2}\sum_{n=1}^{+\infty}
S^{\fuera}_{n}(a)f_n^2\right]+\nonumber\\ & + &
B_{\fuera}\exp\left[-\frac{1}{\textrm{G}} R^{\fuera}_0(a)-
\frac{1}{2}\sum_{n=1}^{+\infty}
R^{\fuera}_{n}(a)f_n^2\right],\nonumber\\ \label{funcionf}
\end{eqnarray}
while in the forbidden region ($a_p<a$), the wave function will be
\begin{eqnarray}
\Psi_{\dentro}(a,f_n)=\exp\left[-\frac{1}{\textrm{G}}
S^{\dentro}_0(a)- \frac{1}{2}\sum_{n=1}^{+\infty}
S^{\dentro}_{n}(a)f_n^2\right]. \label{funciond}
\end{eqnarray}
Where we have considered only the decreasing wave function for the
asymptotic region $a_p<a$ under the barrier.

In the linear regime, around the scale factor $a=a_p$, the wave
function satisfies Eq.~(\ref{wwl}) and can be expressed
as\footnote{ We consider a unique mode $f_n$ for the massive
scalar field as for the case of a positive cosmological constant.
For the general case (multiples modes of the massive scalar field)
the results can be easily generalized.\label{2}}
\begin{eqnarray}
\Psi_p(a,f_n)& = &
\sum_{m_{p}=0}^{+\infty}\left[\gamma_{m_p}Ai(z_p)+\delta_{m_p}Bi(z_p)\right]\times
\nonumber\\ &\times &\exp(-y_{n_p}^2/2){\rm H}_{m_p}(y_{n_p}),
\label{linealizadap}
\end{eqnarray}
where
\begin{eqnarray}
z_p& = &\left(\frac{3\pi}{2\textrm{G}}\right)^{2/3}
\left[\partial_a V_0(a_p)\right]^{1/3}(a-a_{p})+\nonumber\\ & + &
2\left(\frac{2\textrm{G}}{3\pi}\right)^{1/3} \left[\partial_a
V(a_p) \right]^{-2/3}\beta_{n_p},\nonumber\\ \beta_{n_p}& =
&-(2m_p+1)\sqrt{V_n(a_{p})/2}; m_p\in\mathbb{N},\nonumber \\
y_{n_p}& = &\left(2V_n(a_{p})\right)^{1/4}f_n,\; \label{defbetap}
\end{eqnarray}
We match the wave function in the linear regime (see
Eq.~(\ref{linealizadap})) with the wave function under the barrier
$\Psi_{\dentro}(a,f_n)$ (see Eq.~(\ref{funciond})), taking into
account that the background part of $\Psi_{\dentro}(a,f_n)$ is a
decreasing function of the scale factor $a$, whose exponent can be
approximated near $a_p$ by
\begin{equation}
S^{\dentro}_0(a)=\pi\sqrt{\partial_a{V}_0(a_p)}(a-a_p)^{\frac{3}{2}},
\end{equation}
The background part of the FRW wave function outside the barrier
$\Psi_{\fuera}$ in the neighborhood of $a_p$ corresponds to
ingoing and outgoing modes. Therefore $S_0^{\fuera}$ and
$R_0^{\fuera}$ will have the form
\begin{equation}
S_0^{\fuera}(a)=-R_0^{\fuera}(a)=i\pi\sqrt{\partial_a{V_0}(a_p)}(a_p-a)^{\frac{3}{2}}.
\end{equation}
Taking into account that there exists values of the scale factor
close to $a_p$ for which the linear and WBK approximation and the
asymptotic condition $|z_p|\rightarrow+\infty$ hold
simultaneously, we conclude that
\begin{displaymath}
\gamma_{m_p}=\left\{ \begin{array}{ll}
B_{\fuera}(1-k_p)^{l_p}/(2^{2l_p} {l_p}!{ k_{p}}^{l_p+1/2}),
 \; m_p=2l_p, \; l_p\in\mathbb{N},
\\
0, \qquad\qquad\qquad\qquad\qquad\; m_p=2l_p+1, \;
l_p\in\mathbb{N},
\end{array}\right.
\end{displaymath}
\begin{equation}
\delta_{m_p}=0, \; m_p\in\mathbb{N}, \nonumber
\end{equation}
where
\begin{equation}
2k_p=\frac{S^{\dentro}_n(a_p)}{\sqrt{2V_n(a_p)}}+1.
\end{equation}
Matching now the wave function $\Psi_p(a,f_n)$ defined in the
linear regime with $\Psi_{\fuera}(a,f_n)$ corresponding to the
wave function outside the potential barrier we have
\begin{eqnarray}
\Psi_{\fuera}(a,f_n)& \propto &\exp\left[-\frac{1}{\textrm{G}}
S^{\fuera}_0(a)- \frac{1}{2}\sum_{n=1}^{+\infty}
S^{\fuera}_{n}(a)f_n^2\right]-\nonumber\\ & - &
i\exp\left[-\frac{1}{\textrm{G}} R^{\fuera}_0(a)-
\frac{1}{2}\sum_{n=1}^{+\infty}
R^{\fuera}_{n}(a)f_n^2\right],\nonumber \\
\end{eqnarray}
where the proportionality symbol is related to a normalization
constant that we will disregard, and the perturvative parts of the
WBK wave function must satisfy
\begin{eqnarray}
S^{\dentro}_n(a_p)=S^{\fuera}_n(a_p)=R^{\fuera}_n(a_p).
\label{matching}
\end{eqnarray}
On the other hand, similar to the case of positive cosmological
constant, the functions $S^{\dentro}_n(a)$, $R^{\fuera}_n(a)$ and
$R^{\fuera}_n(a)$ have to satisfy regularity conditions which
ensure the good behavior of the wave function on the WBK regime:
\begin{eqnarray}
\mathit{Re}\left[S_n^{\fuera}(a)\right],
\;\mathit{Re}\left[R_n^{\fuera}(a)\right]>0,&& \; \textrm{for} \;
a<a_p,\nonumber\\ \mathit{Re}\left[S_n^{\dentro}(a)\right]>0,&&\;
\textrm{for} \; a_p<a. \label{regularity-}
\end{eqnarray}

\section{Background wave function and matter fluctuations}
\label{V}

As we saw in Sec.~\ref{semiclassical}, the behavior of the wave
function in the WBK regime is determined by background and
perturbative contributions (the matter associated to the vacuum
fluctuations of a massive scalar field conformally coupled to
gravity), which satisfy Eqs.(\ref{sn1},\ref{sn2}).

\subsection{Positive cosmological constant}

In this case, there are two classically allowed regions and a
forbidden one, when the amount of radiation present in the
universe does not excede the maximum of the potential $V(a)$. In
the region $a<a_-$, the ingoing and outgoing background parts of
the wave function, related to $S_0^{\uno}$ and $R_0^{\uno}$, can
be deduced straightforwardly using Eq.~(\ref{sn1}):
\begin{eqnarray}
S_0^{\uno}(a)&=&-R_0^{\uno}(a)=i\frac{3\pi}{2}\int_a^{a_-}\sqrt{-V_0}\;da\nonumber
\\ &=& \frac{i\pi}{2}\frac{1}{\sqrt{\lambda}}\Bigg\{a_+
\left[{\rm E}(\xi_{\uno},\alpha_{\uno})-m{\rm
F}(\xi_{\uno},\alpha_{\uno})\right]-\nonumber\\&-&
\left[\frac{3m+1}{2}-\lambda a^2
\right]\sqrt{\frac{a^2(a_-^2-a^2)}{a_+^2-a^2}}\Bigg\},
\end{eqnarray}
where
\begin{equation}
\xi_{\uno}={\arcsin}\left[\frac{1}{\sqrt{\alpha_{\uno}}}
\left(\frac{a_-^2-a^2}{a_+^2-a^2}\right)^{1/2}\right],\quad
\alpha_{\uno}=\frac{1-m}{1+m}, \nonumber
\end{equation}
and ${\rm F}(\xi_{\uno},\alpha_{\uno})$ and ${\rm
E}(\xi_{\uno},\alpha_{\uno})$ are the elliptic integrals of the
first and second kind respectively \cite{Abramowitz,Gradshteyn}.

In order to study the perturbations, we will now introduce the
Lorentzian conformal time $\bar\eta$ through
\begin{equation}
\frac{dS^{\uno}_0}{da}=-\frac{dR^{\uno}_0}{da}=i\frac{3\pi}{2}\frac{da}{d\bar\eta}.
\nonumber
\end{equation}
The differential equation satisfied by the perturvative parts wave
function, related to $S^{\uno}_n$ and $R^{\uno}_n$, (see
Eq.~(\ref{sn2})) can be linearized introducing the functions
$\nu^{\uno}_n$ defined as
\begin{eqnarray}
S^{\uno}_n(\bar\eta)&=&-i
(\nu^{\uno}_n)'(\bar\eta)/\nu^{\uno}_n(\bar\eta),\nonumber\\
R^{\uno}_n(\bar\eta)&=&-i
(\nu^{\uno}_n)'(-\bar\eta)/\nu^{\uno}_n(-\bar\eta), \label{defnuI}
\end{eqnarray}
where the prime denotes derivative with respect to the Lorentzian
conformal time. In term of the functions $\nu^{\uno}_n$, the
differential equation that satisfy $S^{\uno}_n$ and $R^{\uno}_n$
reduces to
\begin{equation}
(\nu^{\uno}_n)''+\left[(n+1)^2
+\mu^2a^2(\bar\eta)\right]\nu^{\uno}_n=0, \label{linealizarI}
\end{equation}
where $\mu$ is the mass of the scalar field $\Phi$, and
$a(\bar\eta)$ is
\begin{equation}
a(\bar\eta)^2=a_-^2{\rm cd^2}\left [\sqrt{\frac{1+m}{2}}\bar\eta,
\frac{1-m}{1+m}\right], \label{aconfI}
\end{equation}
$\bar\eta \in [-\sqrt{\frac{2}{1+m}}{\rm K}(m),0]$ and ${\rm
cd}\left[x,\frac{1-m}{1+m}\right]$ is a Jacobian elliptic function
\cite{Abramowitz,Gradshteyn}. Eq.~(\ref{linealizarI}) is a
generalized Lam\'{e} differential equation \cite{Ince}. This can be
seen taking into account the explicit expression of the scale
factor $a(\bar\eta)$ given in Eq.~(\ref{aconfI}), the relation
${\rm cd}\left[x+{\rm K}(m),m\right]=-{\rm sn}\left[x,m\right]$
\cite{Abramowitz}, and introducing a new variable
$u\equiv\sqrt{\frac{1+m}{2}}\bar\eta$, so that
Eq.~(\ref{linealizarI}) becomes the generalized Lam\'{e} equation
\begin{equation}
\frac{d^2\nu_n^{\uno}}{du^2}=\left\{N(N+1)k\;{\rm
sn}^2\left[u-{\rm K(k) },k\right]-h\right\}\nu_n^{\uno},
\end{equation}
with
\begin{equation}
N(N+1)=-\frac{\mu^2}{\lambda},\; k=\frac{1-m}{1+m},\; h=2
\frac{(n+1)^2}{1+m}.
\end{equation}

In the classically forbidden region, ($a_-<a<a_+$), the functions
$R_0^{\dos}$ and $S_0^{\dos}$ are
\begin{eqnarray}
S_0^{\dos}(a)& =
&-R_0^{\dos}(a)=-\frac{3\pi}{2}\int_{a_-}^{a}\sqrt{V_0}\;da\nonumber
\\ & = & -\frac{\pi}{2}\frac{1}{\sqrt{\lambda}}\Bigg\{a_+
\left[{\rm E}(\xi_{\dos},\alpha_{\dos})-(1-m){\rm
F}(\xi_{\dos},\alpha_{\dos})\right]+\nonumber\\ &+& \frac{2}{3}
\left[\lambda a^2-1
\right]\sqrt{\frac{(a_+^2-a^2)(a^2-a_-^2)}{a^2}}\Bigg\},
\end{eqnarray}
where
\begin{equation}
\xi_{\dos}={\arcsin}\left[\frac{1}{\sqrt{\alpha_{\dos}}}\left(\frac{a^2-a_-^2}{a^2}\right)^{1/2}\right],\quad
\alpha_{\dos}=\frac{2m}{1+m}. \nonumber
\end{equation}

On the other hand, the differential equation satisfied by the
perturbations $S_n^{\dos}$ and $R_n^{\dos}$ can be simplified as
before:
\begin{equation}
(\nu^{\dos}_n)''-\left[(n+1)^2
+\mu^2a^2(\eta)\right]\nu^{\dos}_n=0, \label{linealizarII}
\end{equation}
where, now, $\eta$ is the Euclidean conformal time related to the
growing and decreasing background terms of the wave function
$\Psi^{\dos}(a,f_n)$ by
\begin{equation}
\frac{dS^{\dos}_0}{da}=-\frac{dR^{\dos}_0}{da}=-\frac{3\pi}{2}\frac{da}{d\eta},
\nonumber
\end{equation}
$a(\eta)$ is the Euclidean scale factor given in
Eq.~(\ref{instanton}), the functions $\nu^{\dos}_n$ are defined as
\begin{eqnarray}
S^{\dos}_n(\eta)&=&-
(\nu^{\dos}_n)'(-\eta)/\nu^{\dos}_n(-\eta),\nonumber\\
R^{\dos}_n(\eta)&=&- (\nu^{\dos}_n)'(\eta)/\nu^{\dos}_n(\eta),
\label{defnuII}
\end{eqnarray}
and the prime denotes derivative with respect to the Euclidean
conformal time $\eta$. As the functions $\nu^{\uno}_n$, their
analogues $\nu_n^{\dos}$, defined under the potential barrier
$V(a)$, satisfy also a generalized Lam\'{e} equation.

Finally, for the asymptotically de Sitter regime ($a>a_+$), the
background parts of the wave function $\Psi_{\tres}(a,f_n)$ are
\begin{eqnarray}
S_0^{\tres}(a)& =
&-R_0^{\tres}(a)=i\frac{3\pi}{2}\int_{a_+}^{a}\sqrt{-V_0}\;da\nonumber
\\ & = & i\frac{\pi}{2}\frac{1}{\sqrt{\lambda}}\Bigg\{a_+
\left[{\rm E}(\xi_{\tres},\alpha_{\tres})-m{\rm
F}(\xi_{\tres},\alpha_{\tres})\right]+\nonumber\\ &+&
\left[\lambda a^2-\frac{3-m}{2}
\right]\sqrt{\frac{(a^2-a_+^2)a^2}{a^2-a_-^2}}\Bigg\},
\end{eqnarray}
where
\begin{equation}
\xi_{\tres}={\arcsin}\left[\frac{a^2-a_+^2}{a^2-a_-^2}\right]^{1/2},\quad
\alpha_{\tres}=\frac{1-m}{1+m}. \nonumber
\end{equation}
The perturbative parts of $\Psi_{\tres}(a,f_n)$ can be obtained
following the same procedure as for $\Psi_{\uno}(a,f_n)$ and
$\Psi_{\dos}(a,f_n)$. We first introduce the Lorentzian time,
$\bar\eta$, by
\begin{equation}
\frac{dS^{\tres}_0}{da}=-\frac{dR^{\tres}_0}{da}=i\frac{3\pi}{2}\frac{da}
{d\bar\eta}, \nonumber
\end{equation}
and define the functions $\nu^{\tres}_n$ by
\begin{eqnarray}
S^{\tres}_n(\bar\eta)&=&-i(\nu^{\tres}_n)'(\bar\eta)/\nu^{\tres}_n(\bar\eta),\nonumber\\
R^{\tres}_n(\bar\eta)&=&-i(\nu^{\tres}_n)'(-\bar\eta)/\nu^{\tres}_n(-\bar\eta),
\label{defnuIII}
\end{eqnarray}
which satisfy
\begin{equation}
(\nu^{\tres}_n)''+\left[(n+1)^2
+\mu^2a^2(\bar\eta)\right]\nu^{\tres}_n=0, \label{linealizarIII}
\end{equation}
where the explicit expression of the scale factor $a(\bar\eta)$ in
the asymptotically de Sitter regime is \cite{Gradshteyn}
\begin{equation}
a(\bar\eta)^2=a_+^2{\rm dc^2}\left[\sqrt{\frac{1+m}{2}}\bar\eta,
\frac{1-m}{1+m}\right], \label{aconfII}
\end{equation}
with $\bar\eta\in [0,\sqrt{\frac{1+m}{2}}{\rm
K}(\frac{1-m}{1+m})]$ and ${\rm dc}\left[x,\frac{1-m}{1+m}\right]$
being a Jacobian elliptic function \cite{Abramowitz,Gradshteyn}.
As before, $\nu_n^{\tres}$ also satisfies a generalized Lam\'{e}
differential equation.

To obtain the explicit expression of the perturbative parts of the
FRW universe, in the case of a positive cosmological constant, it
is necessary to solve the differential equations
(\ref{linealizarI},\ref {linealizarII},\ref{linealizarIII}) that
satisfy the functions $\nu^{\uno}_n$, $\nu^{\dos}_n$ and
$\nu^{\tres}_n$. Nevertheless, it is enough to solve only one of
them since the dependence of the scale factor $a$ on the
Lorentzian time $\bar\eta$ for both classically allowed regions
can be deduced by performing an analytical continuation of $a$,
under the barrier $V(a)$, from the conformal Euclidean $\eta$ to
the Lorentzian one $\bar\eta$. The differential equations
(\ref{linealizarI},\ref{linealizarII},\ref{linealizarIII}) are
related by these analytical continuations and so their solutions
can also be related in the same way. Finally it must be pointed
out that the boundary conditions that $\nu^{\uno}_n$,
$\nu^{\dos}_n$ and $\nu^{\tres}_n$ satisfy are given by the
regularity conditions (\ref{regularity+}) which ensure a good
behavior of the wave function of the universe.

\subsection{Negative cosmological constant}

While for positive cosmological constant, $\lambda$, a FRW
universe filled with radiation can present two classically
disconnected regions, for negative $\lambda$ there is just one
classically allowed region. This section is devote to the latter
case, presenting at the end of our calculations an explicit
example in which the description of the perturbative parts of the
wave function can be carried out analytically. Similarly to the
preceding case, $\lambda>0$, the nonperturbative parts of the wave
function can be obtained from expression (\ref{sn1}). For the
classically allowed region ($a<a_p$), the function $S_n^{\fuera}$
and $R_0^{\fuera}$ related to the background action are
\begin{eqnarray}
S_0^{\fuera} & = & - R_0^{\fuera} =
i\frac{3\pi}{2}\int_{a}^{a_p}\sqrt{-V_0}\;da\nonumber \\ & = & i
\frac{\pi}{2}\left\{\sqrt{m}\left[ -\frac{1+3m}{2\lambda}{\rm
E}(r_{\fuera},s_{\fuera})+\frac{1+m}{\lambda}{\rm
F}(r_{\fuera},s_{\fuera})\right]-\right.\nonumber\\&-&\left.
\sqrt{-\lambda a^2 (a_p^2-a^2)(a^2-a_n^2)}\right\},
\end{eqnarray}
while for the classically forbidden region ($a_p<a$), the action
can be expressed as
\begin{eqnarray}
S_0^{\dentro} & = & \frac{3\pi}{2}\int_{a_p}^{a}\sqrt{V_0}\;
da\nonumber
\\ & = & \frac{\pi}{2}\left[
-\frac{1+3m}{2\lambda}{\rm
E}(r_{\dentro},s_{\dentro})-\frac{1-m}{2\lambda}{\rm
F}(r_{\dentro},s_{\dentro})\right]+\nonumber\\ &+&
\sqrt{-\lambda}\frac{\pi
m}{m+1}(a^2-\frac{1}{\lambda})\sqrt{\frac{(a^2-a_p^2)(a^2-a_n^2)}{a^2}},\nonumber\\
\end{eqnarray}
where $a_n^2= (1+m)/(2\lambda)$ and
\begin{eqnarray}
r_{\fuera}=\arcsin\sqrt{\frac{a_p^2-a^2}{a^2}},\;
s_{\fuera}=\frac{a_p^2}{a_p^2-a_n^2},\nonumber\\
r_{\dentro}=\arcsin\sqrt{\frac{a^2-a_p^2}{a^2}},\;s_{\dentro}=\frac{a_n^2}{a_n^2-a_p^2}.
\end{eqnarray}

The vacuum fluctuations of the massive scaler field, $\Phi$, yield
to perturbative parts, $S_n^{\fuera}$, $R_n^{\fuera}$ and
$S_n^{\dentro}$, in the wave function as described before. The
analogy between the differential equations that govern the
perturbative parts when containing a positive or a negative
cosmological constant $\lambda$ suggests the introduction of the
Lorentzian conformal time $\bar\eta$ and the euclidean one $\eta$
to linearize Eq.~(\ref{sn2}) for the functions $S_n^{\fuera}$,
$R_n^{\fuera}$ and $S_n^{\dentro}$. So, for values of the scale
factor, $a$, smaller than the maximum radius of the collapsing FRW
universe, $a_p$, we define $\bar\eta$ as
\begin{equation}
\frac{dS_0^{\fuera}}{da}=-\frac{dR_0^{\fuera}}{da}=i\frac{3\pi}{2}\frac{da}{d\bar\eta},
\end{equation}
while for $a>a_p$, $\eta$ is given by
\begin{equation}
\frac{dS_0^{\dentro}}{da}=\frac{3\pi}{2}\frac{da}{d\eta}.
\end{equation}
Similarly to the positive $\lambda$ case, we introduce the
functions $\nu_n^{\fuera}$, related to $S_n^{\fuera}$ and
$R_n^{\fuera}$, in the classically allowed region:
\begin{eqnarray}
S^{\fuera}_n(\bar\eta)=-i(\nu^{\fuera}_n)'(\bar\eta)/\nu^{\fuera}_n(\bar\eta),\nonumber\\
R^{\fuera}_n(\bar\eta)=-i(\nu^{\fuera}_n)'(-\bar\eta)/\nu^{\fuera}_n(-\bar\eta),
\label{defnuf}
\end{eqnarray}
where the prime denotes derivative with respect to $\bar\eta$. In
term of the new functions $\nu_n^{\fuera}$, Eq.~(\ref{sn2}) reads
\begin{equation}
(\nu^{\fuera}_n)''+\left[(n+1)^2
+\mu^2a^2(\bar\eta)\right]\nu^{\fuera}_n=0, \label{linealizarf}
\end{equation}
and the explicit expression of the scale factor $a(\bar\eta)$ was
given in Eq.~(\ref{afuera}). For convenience, we rewrite the last
equation in terms of the Weierstrass function
$\mathcal{P}(x|\omega,\omega')$ \cite{Abramowitz,Bateman} as a
generalized lam\'{e} equation Eq.~\cite{Ince}
\begin{equation}
(\nu^{\fuera}_n)''+\left[(n+1)^2+\frac{\mu^2}{3\lambda}
+\frac{\mu^2}{\lambda}\mathcal{P}(\bar\eta|\omega_{\fuera},\omega_{\fuera}')\right]\nu^{\fuera}_n=0,
\label{weief}
\end{equation}
where the so called half periods, $\omega_{\fuera}$ and
$\omega_{\fuera}'$, of the Weierstrass function
$\mathcal{P}(\bar\eta|\omega_{\fuera},\omega'_{\fuera})$ are
\[
\omega_{\fuera}={\rm
K}\left(\frac{m-1}{2m}\right)/\sqrt{m}\;,\quad
\omega_{\fuera}'=i{\rm K}\left(\frac{m+1}{2m}\right)/\sqrt{m}.
\]

Under the potential $V(a)$, Eq.~(\ref{potencial}); i.e. for
$a_p<a$, the wave function must decreases and the unperturbed
Euclidean space-time corresponds to an asymptotically adS wormhole
\cite{Barcelo1}. The linearization of Eq.~(\ref{sn2}) for the
functions $S_n^{\dentro}$ can be made as in the preceding cases,
that is, introducing new functions $\nu_n^{\dentro}$ given by
\begin{eqnarray}
S^{\dentro}_n(\eta)=-(\nu^{\dentro}_n)'(\eta)/\nu^{\dentro}_n(\eta),
\label{defnud}
\end{eqnarray}
where the prime denotes derivative with respect to the conformal
Euclidean time and the functions $\nu_n^{\dentro}$ satisfy
\begin{equation}
(\nu^{\dentro}_n)''-\left[(n+1)^2
+\mu^2a^2(\eta)\right]\nu^{\dentro}_n=0. \label{linealizard}
\end{equation}
The explicit expression of the wormhole scale factor was given in
Eq.~(\ref{wormhole}). In terms of the Weierstrass function, this
equation has the form
\begin{equation}
(\nu^{\dentro}_n)''-\left[(n+1)^2+\frac{\mu^2}{3\lambda}-\frac{\mu^2}{\lambda}
\mathcal{P}(\eta+\omega_{\dentro}|\omega_{\dentro},\omega_{\dentro}')\right]\nu_{n}^{\dentro}=0,
\label{weied}
\end{equation}
where the half periods of the Weierstrass function $\mathcal{P}$
are
\[
\omega_{\dentro}={\rm
K}\left(\frac{1+m}{2m}\right)/{\sqrt{m}},\quad
\omega_{\dentro}'=i{\rm K}\left(\frac{m-1}{2m}\right)/{\sqrt{m}}.
\]

Summarizing, we have presented a method to deal with the behavior
of the wave function of a FRW universe with a cosmological
constant and filled with radiation under the presence of vacuum
fluctuations of a massive scalar field conformally coupled to
gravity in the semiclassical approximation.

\subsection{ An explicit example}

We will illustrate the analysis above by studying the stability of
a radiation-filled FRW universe with negative $\lambda$ in the
case in which the mass of the scalar field is $\mu^2=-2\lambda$,
with $\lambda\ll l_p^{-2}$. This simple choice for the value of
the scalar field mass allows us to solve analytically the
differential equations (\ref{weief},\ref{weied}) for the
perturbations since, in this case the generalized Lam\'{e} equations
reduce to Lam\'{e} equations, whose solutions are known.

Under the potential $V_0(a)$, the perturbative parts of the wave
function, $S_n^{\dentro}$, were expressed in term of the functions
$\nu_n^{\dentro}(\eta)$ in Eq.~(\ref{defnud}) where the functions
$\nu_n^{\dentro}(\eta)$ satisfy a generalized Lam\'{e} differential
equation (see Eq.~(\ref{weied})). In the case under consideration,
$\mu^2=-2\lambda$, this equation becomes
\begin{equation}
(\nu^{\dentro}_n)''-\left[\tilde{h}+N(N+1)
\mathcal{P}(\eta+\omega_{\dentro}|\omega_{\dentro},\omega_{\dentro}')\right]\nu_{n}^{\dentro}=0,
\label{WeierstrassF}
\end{equation}
with $N=1$ and $\tilde{h}= (n+1)^2-\frac{2}{3}$. Since
$N\in\mathbb{N}$ this is a Lam\'{e} equation whose solutions can be
expressed as linear combinations of the linearly independent
solutions $\nu_{1n}^{\dentro}(\eta)$ and
$\nu_{2n}^{\dentro}(\eta)$ given by \cite{Ince}
\begin{eqnarray}
\nu_{1n}^{\dentro}(\eta)=
\frac{\sigma(\eta+\omega_{\dentro}+z_{\dentro}|\omega_{\dentro},\omega_{\dentro}')}
{\sigma(\eta+\omega_{\dentro}|\omega_{\dentro},\omega_{\dentro}')}
\exp\left[-\eta
\zeta(z_{\dentro}|\omega_{\dentro},\omega_{\dentro}')\right],\nonumber\\
\nu_{2n}^{\dentro}(\eta)=
\frac{\sigma(-\eta+\omega_{\dentro}+z_{\dentro}|\omega_{\dentro},\omega_{\dentro}')}
{\sigma(-\eta+\omega_{\dentro}|\omega_{\dentro},\omega_{\dentro}')}
\exp\left[\eta
\zeta(z_{\dentro}|\omega_{\dentro},\omega_{\dentro}')\right],\nonumber\\
\end{eqnarray}
where $\sigma(x|\omega_{\dentro},\omega_{\dentro}')$ and
$\zeta(x|\omega_{\dentro},\omega_{\dentro}')$ are Weierstrass
functions \cite{Abramowitz,Bateman} and the parameter
$z_{\dentro}$ is implicitly defined by
\begin{equation}
\mathcal{P}(z_{\dentro}|\omega_{\dentro},\omega_{\dentro}')=(n+1)^2-\frac{2}{3}\;.
\label{defzd}
\end{equation}

The differential operator that defines Eq.~(\ref{WeierstrassF}) is
a Schr\"{o}dinger operator whose potential is periodic as it can be
expressed in term of the Weierstrass function
$\mathcal{P}(x|\omega_{\dentro},\omega_{\dentro}')$. Therefore,
the solutions will present an infinite number of forbidden and
allowed bands known as Floquet bands and the linear independent
solutions $\nu_{1n}^{\dentro}(\eta)$ and
$\nu_{2n}^{\dentro}(\eta)$ will be characterized by a Floquet
index $\textrm{F}_n^{\dentro}$, independent of $\eta$, such that
\begin{eqnarray}
\nu_{1n}^{\dentro}(\eta+2\omega_{\dentro})& = &
\exp(i\textrm{F}_n^{\dentro})\nu_{1n}^{\dentro}(\eta),\nonumber\\
\nu_{2n}^{\dentro}(\eta+2\omega_{\dentro})& = &
\exp(-i\textrm{F}_n^{\dentro})\nu_{2n}^{\dentro}(\eta).
\label{cuasif}
\end{eqnarray}
So, for the allowed bands, defined by real values of
$\textrm{F}_n^{\dentro}$, the amplitudes of
$\nu_n^{\dentro}(\eta)$, will be in principle bounded from above,
while for the forbidden bands, i.e. for complex values of
$\textrm{F}_n^{\dentro}$, the solutions will be exponentially
increasing or decreasing. In the case under consideration, the
explicit expression of the Floquet index can be deduced using the
following propriety \cite{Bateman}:
$$\sigma(x+2\omega_{\dentro}|\omega_{\dentro},\omega_{\dentro}')=
-\sigma(x|\omega_{\dentro},\omega_{\dentro}')\exp\left[2(x+\omega_{\dentro})\xi_{\dentro}\right]$$
where
$\xi_{\dentro}\equiv\zeta(\omega_{\dentro}|\omega_{\dentro},\omega_{\dentro}')$,
which imply
\begin{equation}
\textrm{F}_n^{\dentro}=2i\left[\omega_{\dentro}\zeta(z_{\dentro}|\omega_{\dentro},\omega_{\dentro}')
-z_{\dentro}\xi_{\dentro}\right].
\end{equation}
As can be seen from this expression, $\textrm{F}_n^{\dentro}$
depends on the parameter $z_{\dentro}$ defined in
Eq.~(\ref{defzd}). So, we have to obtain the possible values of
$z_{\dentro}$ in order to characterize $\textrm{F}_n^{\dentro}$
and the behavior of the linear independent solutions
$\nu_{1n}^{\dentro}(\eta)$ and $\nu_{2n}^{\dentro}(\eta)$.

Since the parameter $z_{\dentro}$ is defined implicitly through
the Weierstrass function
$\mathcal{P}(x|\omega_{\dentro},\omega_{\dentro}')$, we restrict
its values to the fundamental rectangle \cite{Abramowitz}, whose
vertices coincide with the values 0, $\omega_{\dentro}$,
$\omega_{\dentro}'$, and $\omega_{\dentro}+\omega_{\dentro}'$.
That is, $z_{\dentro}$ can belong to the following ranges
\begin{displaymath}
\left\{ \begin{array}{lll} \textrm{Range}\; \mathcal{A}:&
z_{\dentro}=\beta_{\dentro},\;\quad
0<\beta_{\dentro}\leq\omega_{\dentro},\\ \textrm{Range}\;
\mathcal{B}:&
z_{\dentro}=\omega_{\dentro}+i\delta_{\dentro},\;\quad
0\leq\delta_{\dentro}\leq|\omega_{\dentro}'|,\\ \textrm{Range}
\;\mathcal{C}:&
z_{\dentro}=\omega_{\dentro}'+\beta_{\dentro},\;\quad0\leq\beta_{\dentro}\leq\omega_{\dentro},\\
\textrm{Range} \;\mathcal{D}:&
z_{\dentro}=i\delta_{\dentro},\;\quad
0<\delta_{\dentro}\leq|\omega_{\dentro}'|.
\end{array} \right.
\end{displaymath}
When $z_{\dentro}$ takes values in each of the four preceding
ranges, its definition given by Eq.~(\ref{defzd}) implies
\cite{Abramowitz}
\begin{displaymath}
\left\{ \begin{array}{llll} z_{\dentro}\in\textrm{Range}\;
\mathcal{A}&\Longrightarrow &\frac{m+1}{2} \leq(n+1)^2,\\
z_{\dentro}\in\textrm{Range} \;\mathcal{B}&\Longrightarrow& 1 \leq
(n+1)^2 \leq\frac{m+1}{2}, \\ z_{\dentro}\in\textrm{Range}
\;\mathcal{C}&\Longrightarrow& \frac{1-m}{2} \leq(n+1)^2 \leq1, \\
z_{\dentro}\in\textrm{Range}\;\mathcal{D}&\Longrightarrow& (n+1)^2
\leq \frac{1-m}{2}.
\end{array} \right.
\end{displaymath}
Now, remembering that $m>1$ and $n\geq1$, we conclude that the
parameter $z_{\dentro}$ cannot take values in the ranges
$\mathcal{C}$ and $\mathcal{D}$. In the two remaining ranges
$\mathcal{A}$ and $\mathcal{B}$, the Floquet index can be
expressed in term of Theta functions \cite{Bateman}, allowing us
to conclude that the range $\mathcal{A}$ is a forbidden band, i.e.
$\textrm{F}_n^{\dentro}$ is complex, while the range $\mathcal{B}$
is an allowed one, i.e. $\textrm{F}_n^{\dentro}$ is real.

Finally, the general solutions to Eq.~(\ref{WeierstrassF}) will be
a linear combination of $\nu_{1n}^{\dentro}$ and
$\nu_{2n}^{\dentro}$. Using Eq.~(\ref{defnud}), we can write the
functions $S_n^{\dentro}$ in terms of $\nu_{1n}^{\dentro}$ and
$\nu_{2n}^{\dentro}$ with just one free integration constant (this
was expected since $S_n^{\dentro}$ satisfies a first order
differential equation). Therefore, for our purposes and without
loss of generality we can write
\begin{equation}
\nu_{n}^{\dentro}(\eta)=\nu_{1n}^{\dentro}(\eta)+A_{n}^{\dentro}\nu_{2n}^{\dentro}(\eta).
\end{equation}
The constants $A_n^{\dentro}$ have to be chosen so that the
regularity conditions for $S_n^{\dentro}$ hold and the functions
$S_n^{\dentro}$ coincide in the turning point $a_p$ with their
counterparts in the classically allowed region $S_n^{\fuera}$ and
$R_n^{\fuera}$.

The functions $\nu_n^{\fuera}(\bar\eta)$, related to
$S_n^{\fuera}$ and $R_n^{\fuera}$, can be similarly deduced, since
the differential equations (\ref{weief}) also reduce to Lam\'{e}
equations with the following linearly independent solutions
\begin{eqnarray}
\nu_{1n}^{\fuera}(\bar\eta)&=&\frac{\sigma(\bar\eta+\omega_{\fuera}'+z_{\fuera}
|\omega_{\fuera},\omega_{\fuera}')}
{\sigma(\bar\eta+\omega_{\fuera}'
|\omega_{\fuera},\omega_{\fuera}')}
\exp\left[-\bar\eta\zeta(z_{\fuera}|\omega_{\fuera},\omega_{\fuera})\right],\nonumber\\
\nu_{2n}^{\fuera}(\bar\eta)&=&\frac{\sigma(-\bar\eta+\omega_{\fuera}'+z_{\fuera}
|\omega_{\fuera},\omega_{\fuera}')}
{\sigma(-\bar\eta+\omega_{\fuera}'
|\omega_{\fuera},\omega_{\fuera}')}
\exp\left[\bar\eta\zeta(z_{\fuera}|\omega_{\fuera},\omega_{\fuera})\right],\nonumber\\
\end{eqnarray}
where the parameter $z_{\fuera}$ now satisfies the following
relation
\begin{equation}
\mathcal{P}(z_{\fuera}|\omega_{\fuera},\omega_{\fuera}')=-(n+1)^2+\frac{2}{3}.
\end{equation}

These new functions, $\nu_{1n}^{\fuera}(\bar\eta)$ and
$\nu_{2n}^{\fuera}(\bar\eta)$, are quasiperiodic like their
conterparts $\nu_{1n}^{\dentro}(\eta)$ and
$\nu_{2n}^{\dentro}(\eta)$, i.e. they satisfy a relation analogous
to Eq.~(\ref{cuasif}) where the new Floquet index
$\textrm{F}_{n}^{\fuera}$ reads
\begin{equation}
\textrm{F}_{n}^{\fuera}=2i\left[\omega_{\fuera}\zeta(z_{\fuera}|
\omega_{\fuera},\omega'_{\fuera})-z_{\fuera}\xi_{\fuera} \right].
\end{equation}
The values of the parameter $z_{\fuera}$ can be reduced to the
fundamental rectangle of
$\mathcal{P}(x|\omega_{\fuera},\omega'_{\fuera})$, but, as before,
owing to the presence of radiation in the FRW universe ($m>1$) and
the vacuum fluctuations of the massive scalar field ($n\geq1$),
they can only belong to the following ranges:
\begin{displaymath}
\left\{\begin{array}{lll} \textrm{Range}\;\mathcal{E}:
&z_{\fuera}=\omega'_{\fuera}+\beta_{\fuera},\;
0\leq\beta_{\fuera}\leq\omega_{\fuera},\\
\textrm{Range}\;\mathcal{F}:&z_{\fuera}=i\delta_{\fuera},\;
0<\delta_{\fuera}\leq|\omega'_{\fuera}|,
\end{array}\right.
\end{displaymath}
for which
\begin{displaymath}
\left\{ \begin{array}{llll} z_{\fuera}\in\textrm{Range}\;
\mathcal{E}&\Longrightarrow & 1 \leq(n+1)^2 \leq\frac{m+1}{2},\\
z_{\fuera}\in\textrm{Range} \;\mathcal{F}&\Longrightarrow&
\frac{m+1}{2} \leq(n+1)^2.
\end{array} \right.
\end{displaymath}

Note that there is a correspondence between these inequalities,
satisfied in the ranges $\mathcal{E}$, and $\mathcal{F}$ and those
satisfied in the ranges $\mathcal{B}$ and $\mathcal{A}$
respectively under the barrier. This correspondence is due to the
fact that, since the analytic prolongation of $z_{\fuera}$ in
Eq.~(\ref{defzd}) is $z_{\fuera}=iz_{\dentro}$ \cite{Abramowitz},
for each value of $z_{\dentro}$ belonging to the range
$\mathcal{A}$ or $\mathcal{B}$, there is a unique value of
$z_{\fuera}$ belonging to the range $\mathcal{F}$ or $\mathcal{E}$
respectively.

The expression of $F_n^{\fuera}$ in terms of the Theta functions
\cite{Bateman} allow us to conclude that the range $\mathcal{E}$
corresponds to a forbidden band while the range $\mathcal{F}$
corresponds to an allowed one.

Finally, the perturbed part of the wave function, $S_n^{\fuera}$,
can be obtained using Eq.~(\ref{defnuf}) and the solution
\begin{equation}
\nu_{n}^{\fuera}(\bar\eta)=\nu_{1n}^{\fuera}(\bar\eta)+A_n^{\fuera_s}\nu_{2n}^{\fuera}(\bar\eta).
\end{equation}
The function $R_n^{\fuera}$ can be similarly obtained provided
that $\bar\eta$ and $A_n^{\fuera_s}$ are substituted by
$-\bar\eta$ and $A_n^{\fuera_r}$ respectively.

The matching conditions at the turning point $a_p$ require that
$S_n^{\dentro}$, $S_n^{\fuera}$ and $R_n^{\fuera}$ be equal at
this point, so that the independent constant $A_n^{\dentro}$,
$A_n^{\fuera_s}$ and $A_n^{\fuera_r}$ satisfy the relation
\begin{equation}
A_n^{\fuera_s}=A_n^{\fuera_r}=\frac{1-C_n+A_n^{\dentro}(1+C_n)}{1+C_n+A_n^{\dentro}(1-C_n)},
\label{relacionA}
\end{equation}
where
\begin{equation}
C_n=\frac{\omega_{\fuera}}{i\omega_{\dentro}}
\frac{\left[\ln\theta_2(x,q_{\dentro})-\ln\theta_1(x,q_{\dentro})\right]'
     |_{x=\frac{z_{\dentro}}{2\omega_{\dentro}}}}
{\left[\ln\theta_4(x,q_{\fuera})-\ln\theta_1(x,q_{\fuera})\right]'
     |_{x=\frac{z_{\fuera}}{2\omega_{\fuera}}}},
\end{equation}
$q_{\dentro}=\exp[i\pi\omega'_{\dentro}/\omega_{\dentro}]$ and
$q_{\fuera}=\exp[i\pi\omega'_{\fuera}/\omega_{\fuera}]$.

In general, the wave function of the universe, defined in the
whole range of the scale factor, i.e. $a\in[0,+\infty)$, will be a
linear combination of all the wave functions
$\Psi_{A_1^{\dentro},A_2^{\dentro}\ldots}$ defined in
Eqs.~(\ref{funcionf},\ref{funciond}) for the allowed and forbidden
regimes, of the form
\begin{equation}
\Psi=\; \left[\prod_{n\geq1}\int_{\Sigma_n}dA_n^{\dentro}\right]
\zeta_{A_1^{\dentro}A_2^{\dentro}\ldots}\Psi_{A_1^{\dentro},A_2^{\dentro}\ldots}(a,f_n),
\label{superposicion}
\end{equation}
where $\zeta_{A_1^{\dentro}A_2^{\dentro}\ldots}$ are the
coefficient of the linear combinations and $\Sigma_n$ denotes the
set of allowed values of $A_n^{\dentro}$ for each mode, to be
determined.

The wave function $\Psi$ must satisfy a regularity condition that
in terms of the functions $S_n^{\dentro}$, $S_n^{\fuera}$ and
$R_n^{\fuera}$ become a set of inequalities requiring that the
real part of these functions be positive as stated in
Eq.~(\ref{regularity-}). These inequalities select the allowed
values of $A_n^{\dentro}$ for each mode. In others words, the
allowed values of $A_n^{\dentro}$ will be those for which the
minimum, $M[A_n^{\dentro}]$, of $Re[S_n^{\dentro}]$,
$Re[S_n^{\fuera}]$ and $Re[R_n^{\fuera}]$ for all times
($\bar\eta$ and $\eta$) is strictly positive. Naturally, this
minimum will depend both on the amount of radiation present in the
universe and the cosmological constant, which are jointly
represented by $m$, and on the mode $n$ itself. This dependence is
encoded in the parameter $z_{\dentro}$ (and its analytical
continuation $z_{\fuera}$) which may belong either to the range
$\mathcal{A}$ or $\mathcal{B}$.

\begin{figure*}

\includegraphics{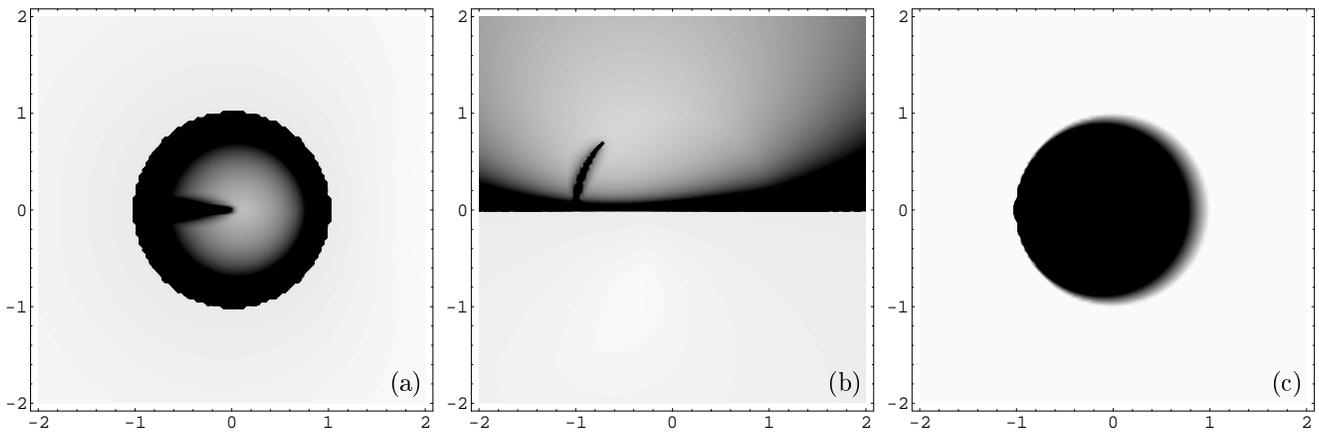}

 \caption{This figure shows the contour plots corresponding to constant values of
$M[A_n^{\dentro}]$ in the complex $A_n^{\dentro}$-plane where
$M[A_n^{\dentro}]$ is defined as the minimum of
$Re[S_n^{\dentro}]$, $Re[S_n^{\fuera}]$ and $Re[R_n^{\fuera}]$ for
all times. In these contour plots, the scale of grey represents
the magnitude of $M$: the darkest (black) one corresponds to
negative infinite values of $M$ while the lightest ones correspond
to positive values and therefore determine the allowed
$A_n^{\dentro}$'s. The parameter $m$ related to the amount of
radiation and the cosmological constant has been set equal to
$m=241$. As discussed in main text, by the regularity conditions
Eq.~(\ref{regularity-}), the allowed values of $A_n^{\dentro}$ are
those for which $M$ is positive. The contour plot in
Fig.~\ref{cp-figura}(a) corresponds to the mode $n=15$. This mode
for the chosen value of $m$ is such that $z_{\dentro}$ belongs to
the range $\mathcal{A}$. The set
$\left\{A_{15}^{\dentro}\in\mathbb{C},1<|A_{15}^{\dentro}|\right\}$
corresponds to the allowed values of $ A_{15}^{\dentro}$ by the
regularity conditions. Fig.~\ref{cp-figura}(b) represents the
contour plot for the mode $n=5$ for which $z_{\dentro}$ belongs to
the range $\mathcal{B}$. The values of $A_5^{\dentro}$ for which
the regularity conditions hold are
$\left\{A_5^{\dentro}\in\mathbb{C},1<|A_5^{\dentro}|\right\}$.
Finally, in Fig.~\ref{cp-figura}(c) shows the contour plot of $M$
for $n=10$. In this case $z_{\dentro}$ lies in the boundary of the
range $\mathcal{A}$ and $\mathcal{B}$. The allowed values of
$A_{10}^{\dentro}$ are
$\left\{A_{10}^{\dentro}\in\mathbb{C},1<|A_{10}^{\dentro}|\right\}$
and $M$ reaches infinite negative values for any other value of
$A_{10}^{\dentro}$.} \label{cp-figura}
\end{figure*}

We have plotted the contours corresponding to constant values of
the minimum $M[A_n^{\dentro}]$ in the complex
$A_n^{\dentro}$-plane for the two possible ranges $\mathcal{A}$
and $\mathcal{B}$ as well as for the value
$z_{\dentro}=\omega_{\dentro}$ that defines the border between
both ranges. These contour plots are shown in
Fig.~\ref{cp-figura}. It can be seen that for $z_{\dentro}\in
\textrm{range}\;\mathcal{A}$, the minimum $M$ is positive outside
the circle of unity radius and centered at the origin, i.e. for
$|A_n^{\dentro}|>1$. Therefore, for fixed amount of radiation
($m=241$ in Fig.~\ref{cp-figura}), the allowed values of
$A_n^{\dentro}$ with $n$ such that
$z_{\dentro}\in\textrm{range}\;\mathcal{A}$ are
$\Sigma_n=\{A_n^{\dentro}\in\mathbb{C}, |A_n^{\dentro}|>1\}$.
Fig.~\ref{cp-figura}-(a) shows the contour plot of $M$ for a
$z_{\dentro}$ in the range $\mathcal{A}$ corresponding to $n=15$
and $m=241$. The modes $n$ such that $z_{\dentro}\in
\textrm{range}\; \mathcal{B}$ exhibit a more complicated behavior.
Depending on the specific mode under study the allowed values
$A_n^{\dentro}$ are either the upper or the lower complex planes,
i.e. $\Sigma_n=\{A_n^{\dentro}\in\mathbb{C},
Im[A_n^{\dentro}]>0\}$ for some $n$ and
$\Sigma_n=\{A_n^{\dentro}\in\mathbb{C}, Im[A_n^{\dentro}]<0\}$ for
the others. Fig.~\ref{cp-figura}-(b) shows the contour plot of $M$
for a $z_{\dentro}$ in the range $\mathcal{B}$ corresponding to
$n=5$ and $m=241$. Other modes will present either a similar plot
or the mirror image with respect to the real axis, as already
discussed. Finally, we see that for those amounts of radiation for
which $m=2(n_B+1)^2-1$ for some $n_B\in\mathbb{N}$, the mode $n_B$
such that $z_{\dentro}$ belongs to both ranges $\mathcal{A}$ and
$\mathcal{B}$, the allowed values of $A_n^{\dentro}$ are
$\Sigma_n=\{A_n^{\dentro}\in\mathbb{C}, |A_n^{\dentro}|>1\}$ as in
Fig.~\ref{cp-figura}-(c). Note however that unlike
Fig.~\ref{cp-figura}-(a), the values $|A_n^{\dentro}|\leq1$
correspond to negative infinite minima $M(A_n^{\dentro})$. In
general, for a given amount of radiation, there will not exist any
such boundary mode and only for very specific fine-tuned amounts
of radiation this will happen.

The existence of a set $\Sigma_n$ for fixed values of the
parameter $m$ is similar to the case studied in \cite{wada}. There
the author constructed the wave function of the gravitons in a de
Sitter background for different boundary conditions. When the
decreasing wave function for the gravitons under the potential
barrier $V(a)$ for $\tilde{K}=0$ was picked up (boundary
conditions similar to the one considered in \cite{Rubakov2}), the
wave function was not uniquely defined or equivalently it can be
constructed as a superposition analogue to
Eq.~(\ref{superposicion}). The situation is rather different when
the increasing wave function of the gravitons under the potential
barrier $V(a)$ is chosen: in this case there is a unique wave
function.

\section{Summary and conclusions}
\label{VI}

In this paper, we have studied the quantum behavior of a
radiation-filled FRW universe with a cosmological constant in the
presence of vacuum fluctuations represented by a massive scalar
field conformally coupled to gravity.

In the semiclassical approximation, the wave function of the
universe can be expressed as linear combinations of outgoing and
ingoing modes in the classically allowed regions and as increasing
and decreasing modes in the classically forbidden ones. For
negative cosmological constant, the matching conditions have been
deduced for natural boundary conditions which pick up the
decreasing wave function in the forbidden region (i.e., in the
asymptotically anti de Sitter Euclidean wormhole). For positive
cosmological constant, the matching conditions have been worked
out for arbitrary boundary conditions and have been applied to the
specific case of the tunneling boundary conditions of the universe
\cite{Vilenkin3}. In this case, the wave function describes a de
Sitter-like universe that contains only outgoing modes in the
asymptotically de Sitter region. These boundary conditions allow
the presence of decreasing and increasing modes under the
potential barrier. However, the ratio between the coefficients of
the increasing and decreasing modes is exponentially suppressed.

Especially important are the regularity conditions that we have
imposed on the wave functions, namely, that they must be finite
and well-behaved everywhere and for every field configurations.
This conditions impose important restriction on the allowed wave
functions as we have shown. In particular, they guarantee that
there are no divergences that could be interpreted as leading to
instabilities of the background configuration (asymptotically anti
de Sitter Euclidean wormhole or asymptotically de Sitter
Lorentzian region, depending on the value of the cosmological
constant). Therefore, we have seen that such regularity conditions
are not empty, at least for some values of the cosmological and
the scalar field mass. Furthermore, we have also shown, in this
case, that they are not too restrictive either. Indeed, there
exist a whole sets of wave functions, characterized by a
continuous index for each mode, which are regular and therefore
feasible candidates for quantum states. In this sense, it is worth
noting that this is true not only in general but also for each
mode separately, i.e, the regularity conditions allow
contributions to the wave function from every single mode without
exception.

With these ingredients, we have obtained explicit solutions for
the background wave function and nonlinear differential equations
that govern the behavior of the vacuum fluctuations. Appropriate
linearization of these equations gives rise to generalized Lam\'{e}
differential equations. As an application of the general
procedures described in this work, we have fully solved the
problem of obtaining the wave function of an asymtotically anti de
Sitter wormhole and its quantum stability against vacuum
fluctuactions represented by conformally coupled scalar field
whose mass is given by $\mu=-2\lambda$. This specific choice has
allowed us to solve the generalized Lam\'{e} equations and thus fully
study the quantum behavior of the vacuum fluctuations. As we have
already discussed, the wormhole boundary conditions and the
regularity condition that the wave function be finite for all
possible values of the scale factor and field configurations
provide the set of allowed quantum wormhole states, which are
therefore stable under vacuum fluctuations. It is worth noting
that the boundary and regularity conditions do not select a single
quantum state as happened in Ref.~\cite{Vilenkin2}, but a set of
allowed quantum states labeled by a continuous parameter for each
mode. This situation is analogous to one of the cases studied in
Ref.~\cite{wada}, where the autor obtained the wave function of
the gravitons in de Sitter space. If this wave function contains
only decreasing modes, the regularity conditions do not select a
unique quantum state for each mode, in opposition to the case
where the boundary condition picked up the increasing wave
function.

In the models considered in Ref.~\cite{wada,Vilenkin2}, as well as
the one studied in this paper the wave functions is well-behaved
in the classically forbidden region, due to quantum gravity
effects, indeed it is not divergent. All these examples show that
in the classically forbidden regions, due to quantum gravity
effects, do not led in principle to infiniteness of the wave
function of the universe and that it is well-behaved in opposition
to the case studied in Ref.~\cite{Rubakov2}, where catastrophic
particle creation led to divergence of the wave function.

\begin{acknowledgments}

M.B.L. is thankful to Alexander Vilenkin for his kindness and
suggesting this work during a visit to Tufts Institute of
Cosmology. M.B.L. is supported by a grant of the Spanish Ministry
of Science and Technology. This work was supported by the DGESIC
under Research Project No. PB97-1218 and PB98-0684.

\end{acknowledgments}

\end{document}